\documentclass[twocolumn,traditabstract,longauth]{aa}
\usepackage{graphicx,amsmath}
\usepackage{epsf}
\usepackage{txfonts}
\usepackage{url}

\usepackage{natbib}
\usepackage{upgreek}

\bibpunct{(}{)}{;}{a}{}{,} % to follow the A&A style

\def\setsymbol#1#2{\expandafter\def\csname #1\endcsname{#2}}
\def\getsymbol#1{\csname #1\endcsname}

\def\Planck{{\it Planck\/}}

\def\allearlypapers{\nocite{planck2011-1.1, planck2011-1.3, planck2011-1.4, planck2011-1.5, planck2011-1.6, planck2011-1.7, planck2011-1.10, planck2011-1.10sup, planck2011-5.1a, planck2011-5.1b, planck2011-5.2a, planck2011-5.2b, planck2011-5.2c, planck2011-6.1, planck2011-6.2, planck2011-6.3a, planck2011-6.4a, planck2011-6.4b, planck2011-6.6, planck2011-7.0, planck2011-7.2, planck2011-7.3, planck2011-7.7a, planck2011-7.7b, planck2011-7.12, planck2011-7.13}}

\newbox\tablebox    \newdimen\tablewidth
\def\leaderfil{\leaders\hbox to 5pt{\hss.\hss}\hfil}
\def\endPlancktable{\tablewidth=\columnwidth 
    $$\hss\copy\tablebox\hss$$
    \vskip-\lastskip\vskip -2pt}

\def\tablenote#1 #2\par{\begingroup \parindent=0.8em
    \abovedisplayshortskip=0pt\belowdisplayshortskip=0pt
    \noindent
    $$\hss\vbox{\hsize\tablewidth \hangindent=\parindent \hangafter=1 \noindent
    \hbox to \parindent{\sup{\rm #1}\hss}\strut#2\strut\par}\hss$$
    \endgroup}

\def\L2{\ifmmode L_2\else $L_2$\fi}

\def\DeltaT{\ifmmode \Delta T\else $\Delta T$\fi}
\def\deltat{\ifmmode \Delta t\else $\Delta t$\fi}
\def\fknee{\ifmmode f_{\rm knee}\else $f_{\rm knee}$\fi}
\def\Fmax{\ifmmode F_{\rm max}\else $F_{\rm max}$\fi}
\def\solar{\ifmmode{\rm M}_{\mathord\odot}\else${\rm M}_{\mathord\odot}$\fi}

\def\inv{\ifmmode^{-1}\else$^{-1}$\fi}
\def\mo{\ifmmode^{-1}\else$^{-1}$\fi}
\def\sup#1{\ifmmode ^{\rm #1}\else $^{\rm #1}$\fi}
\def\expo#1{\ifmmode \times 10^{#1}\else $\times 10^{#1}$\fi}
\def\,{\thinspace}
\def\lsim{\mathrel{\raise .4ex\hbox{\rlap{$<$}\lower 1.2ex\hbox{$\sim$}}}}
\def\gsim{\mathrel{\raise .4ex\hbox{\rlap{$>$}\lower 1.2ex\hbox{$\sim$}}}}

\def\simprop{\mathrel{\raise .4ex\hbox{\rlap{$\propto$}\lower 1.2ex\hbox{$\sim$}}}}
\def\deg{\ifmmode^\circ\else$^\circ$\fi}
\def\pdeg{\ifmmode $\setbox0=\hbox{$^{\circ}$}\rlap{\hskip.11\wd0 .}$^{\circ}
          \else \setbox0=\hbox{$^{\circ}$}\rlap{\hskip.11\wd0 .}$^{\circ}$\fi}
\def\arcs{\ifmmode {^{\scriptstyle\prime\prime}}
          \else $^{\scriptstyle\prime\prime}$\fi}
\def\arcm{\ifmmode {^{\scriptstyle\prime}}
          \else $^{\scriptstyle\prime}$\fi}
\newdimen\sa  \newdimen\sb
\def\parcs{\sa=.07em \sb=.03em
     \ifmmode \hbox{\rlap{.}}^{\scriptstyle\prime\kern -\sb\prime}\hbox{\kern -\sa}
     \else \rlap{.}$^{\scriptstyle\prime\kern -\sb\prime}$\kern -\sa\fi}
\def\parcm{\sa=.08em \sb=.03em
     \ifmmode \hbox{\rlap{.}\kern\sa}^{\scriptstyle\prime}\hbox{\kern-\sb}
     \else \rlap{.}\kern\sa$^{\scriptstyle\prime}$\kern-\sb\fi}
\def\ra[#1 #2 #3.#4]{#1\sup{h}#2\sup{m}#3\sup{s}\llap.#4}
\def\dec[#1 #2 #3.#4]{#1\deg#2\arcm#3\arcs\llap.#4}
\def\deco[#1 #2 #3]{#1\deg#2\arcm#3\arcs}
\def\rra[#1 #2]{#1\sup{h}#2\sup{m}}

\def\dots{\relax\ifmmode \ldots\else $\ldots$\fi}
\def\WHzsr{\ifmmode $W\,Hz\mo\,sr\mo$\else W\,Hz\mo\,sr\mo\fi}
\def\mHz{\ifmmode $\,mHz$\else \,mHz\fi}
\def\GHz{\ifmmode $\,GHz$\else \,GHz\fi}
\def\mKs{\ifmmode $\,mK\,s$^{1/2}\else \,mK\,s$^{1/2}$\fi}
\def\muKs{\ifmmode \,\mu$K\,s$^{1/2}\else \,$\mu$K\,s$^{1/2}$\fi}
\def\muKRJs{\ifmmode \,\mu$K$_{\rm RJ}$\,s$^{1/2}\else \,$\mu$K$_{\rm RJ}$\,s$^{1/2}$\fi}
\def\muKHz{\ifmmode \,\mu$K\,Hz$^{-1/2}\else \,$\mu$K\,Hz$^{-1/2}$\fi}
\def\MJysr{\ifmmode \,$MJy\,sr\mo$\else \,MJy\,sr\mo\fi}
\def\MJysrmK{\ifmmode \,$MJy\,sr\mo$\,mK$_{\rm CMB}\mo\else \,MJy\,sr\mo\,mK$_{\rm CMB}\mo$\fi}
\def\microns{\ifmmode \,\mu$m$\else \,$\mu$m\fi}

\def\muK{\ifmmode \,\mu$K$\else \,$\mu$\hbox{K}\fi}
\def\microK{\ifmmode \,\mu$K$\else \,$\mu$\hbox{K}\fi}
\def\muW{\ifmmode \,\mu$W$\else \,$\mu$\hbox{W}\fi}
\def\kms{\ifmmode $\,km\,s$^{-1}\else \,km\,s$^{-1}$\fi}
\def\kmsMpc{\ifmmode $\,\kms\,Mpc\mo$\else \,\kms\,Mpc\mo\fi}
\setsymbol{LFI:center:frequency:70GHz:units}{70.3\,GHz}
\setsymbol{LFI:center:frequency:44GHz:units}{44.1\,GHz}
\setsymbol{LFI:center:frequency:30GHz:units}{28.5\,GHz}

\setsymbol{LFI:center:frequency:70GHz}{70.3}
\setsymbol{LFI:center:frequency:44GHz}{44.1}
\setsymbol{LFI:center:frequency:30GHz}{28.5}

\setsymbol{LFI:center:frequency:LFI18:Rad:M:units}{71.7\GHz}
\setsymbol{LFI:center:frequency:LFI19:Rad:M:units}{67.5\GHz}
\setsymbol{LFI:center:frequency:LFI20:Rad:M:units}{69.2\GHz}
\setsymbol{LFI:center:frequency:LFI21:Rad:M:units}{70.4\GHz}
\setsymbol{LFI:center:frequency:LFI22:Rad:M:units}{71.5\GHz}
\setsymbol{LFI:center:frequency:LFI23:Rad:M:units}{70.8\GHz}
\setsymbol{LFI:center:frequency:LFI24:Rad:M:units}{44.4\GHz}
\setsymbol{LFI:center:frequency:LFI25:Rad:M:units}{44.0\GHz}
\setsymbol{LFI:center:frequency:LFI26:Rad:M:units}{43.9\GHz}
\setsymbol{LFI:center:frequency:LFI27:Rad:M:units}{28.3\GHz}
\setsymbol{LFI:center:frequency:LFI28:Rad:M:units}{28.8\GHz}
\setsymbol{LFI:center:frequency:LFI18:Rad:S:units}{70.1\GHz}
\setsymbol{LFI:center:frequency:LFI19:Rad:S:units}{69.6\GHz}
\setsymbol{LFI:center:frequency:LFI20:Rad:S:units}{69.5\GHz}
\setsymbol{LFI:center:frequency:LFI21:Rad:S:units}{69.5\GHz}
\setsymbol{LFI:center:frequency:LFI22:Rad:S:units}{72.8\GHz}
\setsymbol{LFI:center:frequency:LFI23:Rad:S:units}{71.3\GHz}
\setsymbol{LFI:center:frequency:LFI24:Rad:S:units}{44.1\GHz}
\setsymbol{LFI:center:frequency:LFI25:Rad:S:units}{44.1\GHz}
\setsymbol{LFI:center:frequency:LFI26:Rad:S:units}{44.1\GHz}
\setsymbol{LFI:center:frequency:LFI27:Rad:S:units}{28.5\GHz}
\setsymbol{LFI:center:frequency:LFI28:Rad:S:units}{28.2\GHz}

\setsymbol{LFI:center:frequency:LFI18:Rad:M}{71.7}
\setsymbol{LFI:center:frequency:LFI19:Rad:M}{67.5}
\setsymbol{LFI:center:frequency:LFI20:Rad:M}{69.2}
\setsymbol{LFI:center:frequency:LFI21:Rad:M}{70.4}
\setsymbol{LFI:center:frequency:LFI22:Rad:M}{71.5}
\setsymbol{LFI:center:frequency:LFI23:Rad:M}{70.8}
\setsymbol{LFI:center:frequency:LFI24:Rad:M}{44.4}
\setsymbol{LFI:center:frequency:LFI25:Rad:M}{44.0}
\setsymbol{LFI:center:frequency:LFI26:Rad:M}{43.9}
\setsymbol{LFI:center:frequency:LFI27:Rad:M}{28.3}
\setsymbol{LFI:center:frequency:LFI28:Rad:M}{28.8}
\setsymbol{LFI:center:frequency:LFI18:Rad:S}{70.1}
\setsymbol{LFI:center:frequency:LFI19:Rad:S}{69.6}
\setsymbol{LFI:center:frequency:LFI20:Rad:S}{69.5}
\setsymbol{LFI:center:frequency:LFI21:Rad:S}{69.5}
\setsymbol{LFI:center:frequency:LFI22:Rad:S}{72.8}
\setsymbol{LFI:center:frequency:LFI23:Rad:S}{71.3}
\setsymbol{LFI:center:frequency:LFI24:Rad:S}{44.1}
\setsymbol{LFI:center:frequency:LFI25:Rad:S}{44.1}
\setsymbol{LFI:center:frequency:LFI26:Rad:S}{44.1}
\setsymbol{LFI:center:frequency:LFI27:Rad:S}{28.5}
\setsymbol{LFI:center:frequency:LFI28:Rad:S}{28.2}

\setsymbol{LFI:white:noise:sensitivity:70GHz:units}{152.6\muKs}
\setsymbol{LFI:white:noise:sensitivity:44GHz:units}{173.1\muKs}
\setsymbol{LFI:white:noise:sensitivity:30GHz:units}{146.8\muKs}

\setsymbol{LFI:white:noise:sensitivity:70GHz}{152.6}
\setsymbol{LFI:white:noise:sensitivity:44GHz}{173.1}
\setsymbol{LFI:white:noise:sensitivity:30GHz}{146.8}

\setsymbol{LFI:white:noise:sensitivity:LFI18:Rad:M:units}{512.0\muKs}
\setsymbol{LFI:white:noise:sensitivity:LFI19:Rad:M:units}{581.4\muKs}
\setsymbol{LFI:white:noise:sensitivity:LFI20:Rad:M:units}{590.8\muKs}
\setsymbol{LFI:white:noise:sensitivity:LFI21:Rad:M:units}{455.2\muKs}
\setsymbol{LFI:white:noise:sensitivity:LFI22:Rad:M:units}{492.0\muKs}
\setsymbol{LFI:white:noise:sensitivity:LFI23:Rad:M:units}{507.7\muKs}
\setsymbol{LFI:white:noise:sensitivity:LFI24:Rad:M:units}{462.2\muKs}
\setsymbol{LFI:white:noise:sensitivity:LFI25:Rad:M:units}{413.6\muKs}
\setsymbol{LFI:white:noise:sensitivity:LFI26:Rad:M:units}{478.6\muKs}
\setsymbol{LFI:white:noise:sensitivity:LFI27:Rad:M:units}{277.7\muKs}
\setsymbol{LFI:white:noise:sensitivity:LFI28:Rad:M:units}{312.3\muKs}
\setsymbol{LFI:white:noise:sensitivity:LFI18:Rad:S:units}{465.7\muKs}
\setsymbol{LFI:white:noise:sensitivity:LFI19:Rad:S:units}{555.6\muKs}
\setsymbol{LFI:white:noise:sensitivity:LFI20:Rad:S:units}{623.2\muKs}
\setsymbol{LFI:white:noise:sensitivity:LFI21:Rad:S:units}{564.1\muKs}
\setsymbol{LFI:white:noise:sensitivity:LFI22:Rad:S:units}{534.4\muKs}
\setsymbol{LFI:white:noise:sensitivity:LFI23:Rad:S:units}{542.4\muKs}
\setsymbol{LFI:white:noise:sensitivity:LFI24:Rad:S:units}{399.2\muKs}
\setsymbol{LFI:white:noise:sensitivity:LFI25:Rad:S:units}{392.6\muKs}
\setsymbol{LFI:white:noise:sensitivity:LFI26:Rad:S:units}{418.6\muKs}
\setsymbol{LFI:white:noise:sensitivity:LFI27:Rad:S:units}{302.9\muKs}
\setsymbol{LFI:white:noise:sensitivity:LFI28:Rad:S:units}{285.3\muKs}

\setsymbol{LFI:white:noise:sensitivity:LFI18:Rad:M}{512.0}
\setsymbol{LFI:white:noise:sensitivity:LFI19:Rad:M}{581.4}
\setsymbol{LFI:white:noise:sensitivity:LFI20:Rad:M}{590.8}
\setsymbol{LFI:white:noise:sensitivity:LFI21:Rad:M}{455.2}
\setsymbol{LFI:white:noise:sensitivity:LFI22:Rad:M}{492.0}
\setsymbol{LFI:white:noise:sensitivity:LFI23:Rad:M}{507.7}
\setsymbol{LFI:white:noise:sensitivity:LFI24:Rad:M}{462.2}
\setsymbol{LFI:white:noise:sensitivity:LFI25:Rad:M}{413.6}
\setsymbol{LFI:white:noise:sensitivity:LFI26:Rad:M}{478.6}
\setsymbol{LFI:white:noise:sensitivity:LFI27:Rad:M}{277.7}
\setsymbol{LFI:white:noise:sensitivity:LFI28:Rad:M}{312.3}
\setsymbol{LFI:white:noise:sensitivity:LFI18:Rad:S}{465.7}
\setsymbol{LFI:white:noise:sensitivity:LFI19:Rad:S}{555.6}
\setsymbol{LFI:white:noise:sensitivity:LFI20:Rad:S}{623.2}
\setsymbol{LFI:white:noise:sensitivity:LFI21:Rad:S}{564.1}
\setsymbol{LFI:white:noise:sensitivity:LFI22:Rad:S}{534.4}
\setsymbol{LFI:white:noise:sensitivity:LFI23:Rad:S}{542.4}
\setsymbol{LFI:white:noise:sensitivity:LFI24:Rad:S}{399.2}
\setsymbol{LFI:white:noise:sensitivity:LFI25:Rad:S}{392.6}
\setsymbol{LFI:white:noise:sensitivity:LFI26:Rad:S}{418.6}
\setsymbol{LFI:white:noise:sensitivity:LFI27:Rad:S}{302.9}
\setsymbol{LFI:white:noise:sensitivity:LFI28:Rad:S}{285.3}

\setsymbol{LFI:knee:frequency:70GHz:units}{29.5\mHz}
\setsymbol{LFI:knee:frequency:44GHz:units}{56.2\mHz}
\setsymbol{LFI:knee:frequency:30GHz:units}{113.7\mHz}

\setsymbol{LFI:knee:frequency:70GHz}{29.5}
\setsymbol{LFI:knee:frequency:44GHz}{56.2}
\setsymbol{LFI:knee:frequency:30GHz}{113.7}

\setsymbol{LFI:knee:frequency:LFI18:Rad:M:units}{16.3\mHz}
\setsymbol{LFI:knee:frequency:LFI19:Rad:M:units}{15.1\mHz}
\setsymbol{LFI:knee:frequency:LFI20:Rad:M:units}{18.7\mHz}
\setsymbol{LFI:knee:frequency:LFI21:Rad:M:units}{37.2\mHz}
\setsymbol{LFI:knee:frequency:LFI22:Rad:M:units}{12.7\mHz}
\setsymbol{LFI:knee:frequency:LFI23:Rad:M:units}{34.6\mHz}
\setsymbol{LFI:knee:frequency:LFI24:Rad:M:units}{46.2\mHz}
\setsymbol{LFI:knee:frequency:LFI25:Rad:M:units}{24.9\mHz}
\setsymbol{LFI:knee:frequency:LFI26:Rad:M:units}{67.6\mHz}
\setsymbol{LFI:knee:frequency:LFI27:Rad:M:units}{187.4\mHz}
\setsymbol{LFI:knee:frequency:LFI28:Rad:M:units}{122.2\mHz}
\setsymbol{LFI:knee:frequency:LFI18:Rad:S:units}{17.7\mHz}
\setsymbol{LFI:knee:frequency:LFI19:Rad:S:units}{22.0\mHz}
\setsymbol{LFI:knee:frequency:LFI20:Rad:S:units}{8.7\mHz}
\setsymbol{LFI:knee:frequency:LFI21:Rad:S:units}{25.9\mHz}
\setsymbol{LFI:knee:frequency:LFI22:Rad:S:units}{15.8\mHz}
\setsymbol{LFI:knee:frequency:LFI23:Rad:S:units}{129.8\mHz}
\setsymbol{LFI:knee:frequency:LFI24:Rad:S:units}{100.9\mHz}
\setsymbol{LFI:knee:frequency:LFI25:Rad:S:units}{38.9\mHz}
\setsymbol{LFI:knee:frequency:LFI26:Rad:S:units}{58.9\mHz}
\setsymbol{LFI:knee:frequency:LFI27:Rad:S:units}{104.4\mHz}
\setsymbol{LFI:knee:frequency:LFI28:Rad:S:units}{40.7\mHz}

\setsymbol{LFI:knee:frequency:LFI18:Rad:M}{16.3}
\setsymbol{LFI:knee:frequency:LFI19:Rad:M}{15.1}
\setsymbol{LFI:knee:frequency:LFI20:Rad:M}{18.7}
\setsymbol{LFI:knee:frequency:LFI21:Rad:M}{37.2}
\setsymbol{LFI:knee:frequency:LFI22:Rad:M}{12.7}
\setsymbol{LFI:knee:frequency:LFI23:Rad:M}{34.6}
\setsymbol{LFI:knee:frequency:LFI24:Rad:M}{46.2}
\setsymbol{LFI:knee:frequency:LFI25:Rad:M}{24.9}
\setsymbol{LFI:knee:frequency:LFI26:Rad:M}{67.6}
\setsymbol{LFI:knee:frequency:LFI27:Rad:M}{187.4}
\setsymbol{LFI:knee:frequency:LFI28:Rad:M}{122.2}
\setsymbol{LFI:knee:frequency:LFI18:Rad:S}{17.7}
\setsymbol{LFI:knee:frequency:LFI19:Rad:S}{22.0}
\setsymbol{LFI:knee:frequency:LFI20:Rad:S}{8.7}
\setsymbol{LFI:knee:frequency:LFI21:Rad:S}{25.9}
\setsymbol{LFI:knee:frequency:LFI22:Rad:S}{15.8}
\setsymbol{LFI:knee:frequency:LFI23:Rad:S}{129.8}
\setsymbol{LFI:knee:frequency:LFI24:Rad:S}{100.9}
\setsymbol{LFI:knee:frequency:LFI25:Rad:S}{38.9}
\setsymbol{LFI:knee:frequency:LFI26:Rad:S}{58.9}
\setsymbol{LFI:knee:frequency:LFI27:Rad:S}{104.4}
\setsymbol{LFI:knee:frequency:LFI28:Rad:S}{40.7}

\setsymbol{LFI:slope:70GHz:units}{$-1.03$\mHz}
\setsymbol{LFI:slope:44GHz:units}{$-0.89$\mHz}
\setsymbol{LFI:slope:30GHz:units}{$-0.87$\mHz}

\setsymbol{LFI:slope:70GHz}{$-1.03$}
\setsymbol{LFI:slope:44GHz}{$-0.89$}
\setsymbol{LFI:slope:30GHz}{$-0.87$}

\setsymbol{LFI:slope:LFI18:Rad:M:units}{$-1.04$\mHz}
\setsymbol{LFI:slope:LFI19:Rad:M:units}{$-1.09$\mHz}
\setsymbol{LFI:slope:LFI20:Rad:M:units}{$-0.69$\mHz}
\setsymbol{LFI:slope:LFI21:Rad:M:units}{$-1.56$\mHz}
\setsymbol{LFI:slope:LFI22:Rad:M:units}{$-1.01$\mHz}
\setsymbol{LFI:slope:LFI23:Rad:M:units}{$-0.96$\mHz}
\setsymbol{LFI:slope:LFI24:Rad:M:units}{$-0.83$\mHz}
\setsymbol{LFI:slope:LFI25:Rad:M:units}{$-0.91$\mHz}
\setsymbol{LFI:slope:LFI26:Rad:M:units}{$-0.95$\mHz}
\setsymbol{LFI:slope:LFI27:Rad:M:units}{$-0.87$\mHz}
\setsymbol{LFI:slope:LFI28:Rad:M:units}{$-0.88$\mHz}
\setsymbol{LFI:slope:LFI18:Rad:S:units}{$-1.15$\mHz}
\setsymbol{LFI:slope:LFI19:Rad:S:units}{$-1.00$\mHz}
\setsymbol{LFI:slope:LFI20:Rad:S:units}{$-0.95$\mHz}
\setsymbol{LFI:slope:LFI21:Rad:S:units}{$-0.92$\mHz}
\setsymbol{LFI:slope:LFI22:Rad:S:units}{$-1.01$\mHz}
\setsymbol{LFI:slope:LFI23:Rad:S:units}{$-0.95$\mHz}
\setsymbol{LFI:slope:LFI24:Rad:S:units}{$-0.73$\mHz}
\setsymbol{LFI:slope:LFI25:Rad:S:units}{$-1.16$\mHz}
\setsymbol{LFI:slope:LFI26:Rad:S:units}{$-0.79$\mHz}
\setsymbol{LFI:slope:LFI27:Rad:S:units}{$-0.82$\mHz}
\setsymbol{LFI:slope:LFI28:Rad:S:units}{$-0.91$\mHz}

\setsymbol{LFI:slope:LFI18:Rad:M}{$-1.04$}
\setsymbol{LFI:slope:LFI19:Rad:M}{$-1.09$}
\setsymbol{LFI:slope:LFI20:Rad:M}{$-0.69$}
\setsymbol{LFI:slope:LFI21:Rad:M}{$-1.56$}
\setsymbol{LFI:slope:LFI22:Rad:M}{$-1.01$}
\setsymbol{LFI:slope:LFI23:Rad:M}{$-0.96$}
\setsymbol{LFI:slope:LFI24:Rad:M}{$-0.83$}
\setsymbol{LFI:slope:LFI25:Rad:M}{$-0.91$}
\setsymbol{LFI:slope:LFI26:Rad:M}{$-0.95$}
\setsymbol{LFI:slope:LFI27:Rad:M}{$-0.87$}
\setsymbol{LFI:slope:LFI28:Rad:M}{$-0.88$}
\setsymbol{LFI:slope:LFI18:Rad:S}{$-1.15$}
\setsymbol{LFI:slope:LFI19:Rad:S}{$-1.00$}
\setsymbol{LFI:slope:LFI20:Rad:S}{$-0.95$}
\setsymbol{LFI:slope:LFI21:Rad:S}{$-0.92$}
\setsymbol{LFI:slope:LFI22:Rad:S}{$-1.01$}
\setsymbol{LFI:slope:LFI23:Rad:S}{$-0.95$}
\setsymbol{LFI:slope:LFI24:Rad:S}{$-0.73$}
\setsymbol{LFI:slope:LFI25:Rad:S}{$-1.16$}
\setsymbol{LFI:slope:LFI26:Rad:S}{$-0.79$}
\setsymbol{LFI:slope:LFI27:Rad:S}{$-0.82$}
\setsymbol{LFI:slope:LFI28:Rad:S}{$-0.91$}

\setsymbol{LFI:FWHM:70GHz:units}{13\parcm01}
\setsymbol{LFI:FWHM:44GHz:units}{27\parcm92}
\setsymbol{LFI:FWHM:30GHz:units}{32\parcm65}

\setsymbol{LFI:FWHM:70GHz}{13.01}
\setsymbol{LFI:FWHM:44GHz}{27.92}
\setsymbol{LFI:FWHM:30GHz}{32.65}

\setsymbol{LFI:FWHM:LFI18:units}{13\parcm39}
\setsymbol{LFI:FWHM:LFI19:units}{13\parcm01}
\setsymbol{LFI:FWHM:LFI20:units}{12\parcm75}
\setsymbol{LFI:FWHM:LFI21:units}{12\parcm74}
\setsymbol{LFI:FWHM:LFI22:units}{12\parcm87}
\setsymbol{LFI:FWHM:LFI23:units}{13\parcm27}
\setsymbol{LFI:FWHM:LFI24:units}{22\parcm98}
\setsymbol{LFI:FWHM:LFI25:units}{30\parcm46}
\setsymbol{LFI:FWHM:LFI26:units}{30\parcm31}
\setsymbol{LFI:FWHM:LFI27:units}{32\parcm65}
\setsymbol{LFI:FWHM:LFI28:units}{32\parcm66}

\setsymbol{LFI:FWHM:LFI18}{13.39}
\setsymbol{LFI:FWHM:LFI19}{13.01}
\setsymbol{LFI:FWHM:LFI20}{12.75}
\setsymbol{LFI:FWHM:LFI21}{12.74}
\setsymbol{LFI:FWHM:LFI22}{12.87}
\setsymbol{LFI:FWHM:LFI23}{13.27}
\setsymbol{LFI:FWHM:LFI24}{22.98}
\setsymbol{LFI:FWHM:LFI25}{30.46}
\setsymbol{LFI:FWHM:LFI26}{30.31}
\setsymbol{LFI:FWHM:LFI27}{32.65}
\setsymbol{LFI:FWHM:LFI28}{32.66}

\setsymbol{LFI:FWHM:uncertainty:LFI18:units}{0.170\arcm}
\setsymbol{LFI:FWHM:uncertainty:LFI19:units}{0.174\arcm}
\setsymbol{LFI:FWHM:uncertainty:LFI20:units}{0.170\arcm}
\setsymbol{LFI:FWHM:uncertainty:LFI21:units}{0.156\arcm}
\setsymbol{LFI:FWHM:uncertainty:LFI22:units}{0.164\arcm}
\setsymbol{LFI:FWHM:uncertainty:LFI23:units}{0.171\arcm}
\setsymbol{LFI:FWHM:uncertainty:LFI24:units}{0.652\arcm}
\setsymbol{LFI:FWHM:uncertainty:LFI25:units}{1.075\arcm}
\setsymbol{LFI:FWHM:uncertainty:LFI26:units}{1.131\arcm}
\setsymbol{LFI:FWHM:uncertainty:LFI27:units}{1.266\arcm}
\setsymbol{LFI:FWHM:uncertainty:LFI28:units}{1.287\arcm}

\setsymbol{LFI:FWHM:uncertainty:LFI18}{0.170}
\setsymbol{LFI:FWHM:uncertainty:LFI19}{0.174}
\setsymbol{LFI:FWHM:uncertainty:LFI20}{0.170}
\setsymbol{LFI:FWHM:uncertainty:LFI21}{0.156}
\setsymbol{LFI:FWHM:uncertainty:LFI22}{0.164}
\setsymbol{LFI:FWHM:uncertainty:LFI23}{0.171}
\setsymbol{LFI:FWHM:uncertainty:LFI24}{0.652}
\setsymbol{LFI:FWHM:uncertainty:LFI25}{1.075}
\setsymbol{LFI:FWHM:uncertainty:LFI26}{1.131}
\setsymbol{LFI:FWHM:uncertainty:LFI27}{1.266}
\setsymbol{LFI:FWHM:uncertainty:LFI28}{1.287}

\setsymbol{HFI:center:frequency:100GHz:units}{100\,GHz}
\setsymbol{HFI:center:frequency:143GHz:units}{143\,GHz}
\setsymbol{HFI:center:frequency:217GHz:units}{217\,GHz}
\setsymbol{HFI:center:frequency:353GHz:units}{353\,GHz}
\setsymbol{HFI:center:frequency:545GHz:units}{545\,GHz}
\setsymbol{HFI:center:frequency:857GHz:units}{857\,GHz}

\setsymbol{HFI:center:frequency:100GHz}{100}
\setsymbol{HFI:center:frequency:143GHz}{143}
\setsymbol{HFI:center:frequency:217GHz}{217}
\setsymbol{HFI:center:frequency:353GHz}{353}
\setsymbol{HFI:center:frequency:545GHz}{545}
\setsymbol{HFI:center:frequency:857GHz}{857}

\setsymbol{HFI:Ndetectors:100GHz}{8}
\setsymbol{HFI:Ndetectors:143GHz}{11}
\setsymbol{HFI:Ndetectors:217GHz}{12}
\setsymbol{HFI:Ndetectors:353GHz}{12}
\setsymbol{HFI:Ndetectors:545GHz}{3}
\setsymbol{HFI:Ndetectors:857GHz}{4}

\setsymbol{HFI:FWHM:Maps:100GHz:units}{9\parcm88}
\setsymbol{HFI:FWHM:Maps:143GHz:units}{7\parcm18}
\setsymbol{HFI:FWHM:Maps:217GHz:units}{4\parcm87}
\setsymbol{HFI:FWHM:Maps:353GHz:units}{4\parcm65}
\setsymbol{HFI:FWHM:Maps:545GHz:units}{4\parcm72}
\setsymbol{HFI:FWHM:Maps:857GHz:units}{4\parcm39}
\setsymbol{HFI:FWHM:Maps:100GHz}{9.88}
\setsymbol{HFI:FWHM:Maps:143GHz}{7.18}
\setsymbol{HFI:FWHM:Maps:217GHz}{4.87}
\setsymbol{HFI:FWHM:Maps:353GHz}{4.65}
\setsymbol{HFI:FWHM:Maps:545GHz}{4.72}
\setsymbol{HFI:FWHM:Maps:857GHz}{4.39}

\setsymbol{HFI:beam:ellipticity:Maps:100GHz}{1.15}
\setsymbol{HFI:beam:ellipticity:Maps:143GHz}{1.01}
\setsymbol{HFI:beam:ellipticity:Maps:217GHz}{1.06}
\setsymbol{HFI:beam:ellipticity:Maps:353GHz}{1.05}
\setsymbol{HFI:beam:ellipticity:Maps:545GHz}{1.14}
\setsymbol{HFI:beam:ellipticity:Maps:857GHz}{1.19}

\setsymbol{HFI:FWHM:Mars:100GHz:units}{9\parcm37}
\setsymbol{HFI:FWHM:Mars:143GHz:units}{7\parcm04}
\setsymbol{HFI:FWHM:Mars:217GHz:units}{4\parcm68}
\setsymbol{HFI:FWHM:Mars:353GHz:units}{4\parcm43}
\setsymbol{HFI:FWHM:Mars:545GHz:units}{3\parcm80}
\setsymbol{HFI:FWHM:Mars:857GHz:units}{3\parcm67}

\setsymbol{HFI:FWHM:Mars:100GHz}{9.37}
\setsymbol{HFI:FWHM:Mars:143GHz}{7.04}
\setsymbol{HFI:FWHM:Mars:217GHz}{4.68}
\setsymbol{HFI:FWHM:Mars:353GHz}{4.43}
\setsymbol{HFI:FWHM:Mars:545GHz}{3.80}
\setsymbol{HFI:FWHM:Mars:857GHz}{3.67}

\setsymbol{HFI:beam:ellipticity:Mars:100GHz}{1.18}
\setsymbol{HFI:beam:ellipticity:Mars:143GHz}{1.03}
\setsymbol{HFI:beam:ellipticity:Mars:217GHz}{1.14}
\setsymbol{HFI:beam:ellipticity:Mars:353GHz}{1.09}
\setsymbol{HFI:beam:ellipticity:Mars:545GHz}{1.25}
\setsymbol{HFI:beam:ellipticity:Mars:857GHz}{1.03}

\setsymbol{HFI:CMB:relative:calibration:100GHz}{$\lsim 1\%$}
\setsymbol{HFI:CMB:relative:calibration:143GHz}{$\lsim 1\%$}
\setsymbol{HFI:CMB:relative:calibration:217GHz}{$\lsim 1\%$}
\setsymbol{HFI:CMB:relative:calibration:353GHz}{$\lsim 1\%$}
\setsymbol{HFI:CMB:relative:calibration:545GHz}{}
\setsymbol{HFI:CMB:relative:calibration:857GHz}{}

\setsymbol{HFI:CMB:absolute:calibration:100GHz}{$\lsim 2\%$}
\setsymbol{HFI:CMB:absolute:calibration:143GHz}{$\lsim 2\%$}
\setsymbol{HFI:CMB:absolute:calibration:217GHz}{$\lsim 2\%$}
\setsymbol{HFI:CMB:absolute:calibration:353GHz}{$\lsim 2\%$}
\setsymbol{HFI:CMB:absolute:calibration:545GHz}{}
\setsymbol{HFI:CMB:absolute:calibration:857GHz}{}

\setsymbol{HFI:FIRAS:gain:calibration:accuracy:statistical:100GHz}{}
\setsymbol{HFI:FIRAS:gain:calibration:accuracy:statistical:143GHz}{}
\setsymbol{HFI:FIRAS:gain:calibration:accuracy:statistical:217GHz}{}
\setsymbol{HFI:FIRAS:gain:calibration:accuracy:statistical:353GHz}{2.5\%}
\setsymbol{HFI:FIRAS:gain:calibration:accuracy:statistical:545GHz}{1\%}
\setsymbol{HFI:FIRAS:gain:calibration:accuracy:statistical:857GHz}{0.5\%}

\setsymbol{HFI:FIRAS:gain:calibration:accuracy:systematic:100GHz}{}
\setsymbol{HFI:FIRAS:gain:calibration:accuracy:systematic:143GHz}{}
\setsymbol{HFI:FIRAS:gain:calibration:accuracy:systematic:217GHz}{}
\setsymbol{HFI:FIRAS:gain:calibration:accuracy:systematic:353GHz}{}
\setsymbol{HFI:FIRAS:gain:calibration:accuracy:systematic:545GHz}{7\%}
\setsymbol{HFI:FIRAS:gain:calibration:accuracy:systematic:857GHz}{7\%}

\setsymbol{HFI:FIRAS:zero:point:accuracy:100GHz:units}{0.8\MJysr}
\setsymbol{HFI:FIRAS:zero:point:accuracy:143GHz:units}{}
\setsymbol{HFI:FIRAS:zero:point:accuracy:217GHz:units}{}
\setsymbol{HFI:FIRAS:zero:point:accuracy:353GHz:units}{1.4\MJysr}
\setsymbol{HFI:FIRAS:zero:point:accuracy:545GHz:units}{2.2\MJysr}
\setsymbol{HFI:FIRAS:zero:point:accuracy:857GHz:units}{1.7\MJysr}

\setsymbol{HFI:FIRAS:zero:point:accuracy:100GHz}{0.8}
\setsymbol{HFI:FIRAS:zero:point:accuracy:143GHz}{}
\setsymbol{HFI:FIRAS:zero:point:accuracy:217GHz}{}
\setsymbol{HFI:FIRAS:zero:point:accuracy:353GHz}{1.4}
\setsymbol{HFI:FIRAS:zero:point:accuracy:545GHz}{2.2}
\setsymbol{HFI:FIRAS:zero:point:accuracy:857GHz}{1.7}

\setsymbol{HFI:unit:conversion:100GHz:units}{0.2415\MJysrmK}
\setsymbol{HFI:unit:conversion:143GHz:units}{0.3694\MJysrmK}
\setsymbol{HFI:unit:conversion:217GHz:units}{0.4811\MJysrmK}
\setsymbol{HFI:unit:conversion:353GHz:units}{0.2883\MJysrmK}
\setsymbol{HFI:unit:conversion:545GHz:units}{0.05826\MJysrmK}
\setsymbol{HFI:unit:conversion:857GHz:units}{0.002238\MJysrmK}

\setsymbol{HFI:unit:conversion:100GHz}{0.2415}
\setsymbol{HFI:unit:conversion:143GHz}{0.3694}
\setsymbol{HFI:unit:conversion:217GHz}{0.4811}
\setsymbol{HFI:unit:conversion:353GHz}{0.2883}
\setsymbol{HFI:unit:conversion:545GHz}{0.05826}
\setsymbol{HFI:unit:conversion:857GHz}{0.002238}

\setsymbol{HFI:colour:correction:alpha=-2:V1.01:100GHz}{0.9893}
\setsymbol{HFI:colour:correction:alpha=-2:V1.01:143GHz}{0.9759}
\setsymbol{HFI:colour:correction:alpha=-2:V1.01:217GHz}{1.0007}
\setsymbol{HFI:colour:correction:alpha=-2:V1.01:353GHz}{1.0028}
\setsymbol{HFI:colour:correction:alpha=-2:V1.01:545GHz}{1.0019}
\setsymbol{HFI:colour:correction:alpha=-2:V1.01:857GHz}{0.9889}

\setsymbol{HFI:colour:correction:alpha=0:V1.01:100GHz}{1.0008}
\setsymbol{HFI:colour:correction:alpha=0:V1.01:143GHz}{1.0148}
\setsymbol{HFI:colour:correction:alpha=0:V1.01:217GHz}{0.9909}
\setsymbol{HFI:colour:correction:alpha=0:V1.01:353GHz}{0.9888}
\setsymbol{HFI:colour:correction:alpha=0:V1.01:545GHz}{0.9878}
\setsymbol{HFI:colour:correction:alpha=0:V1.01:857GHz}{1.0014}

\newcommand{\planck}{\textit{Planck\/}}  % for \planck rather than \Planck

\begin{document}

\title{\planck\ Early Results. XX. New light on anomalous microwave emission from spinning dust grains}
%This author list corresponds to \title{Author list for Proj. Ref. 7.2: New light on anomalous microwave emission from spinning dust grains}
%Prepared by R. Leonardi (rleonardi@sciops.esa.int), ESAC/ESA, on 10MAY2011
%This version is from 16 May 2011 at 12:00 CET
%\subtitle{There are 215 co-authors in this list}
\author{\small
Planck Collaboration:
P.~A.~R.~Ade\inst{72}
\and
N.~Aghanim\inst{46}
\and
M.~Arnaud\inst{58}
\and
M.~Ashdown\inst{56, 4}
\and
J.~Aumont\inst{46}
\and
C.~Baccigalupi\inst{70}
\and
A.~Balbi\inst{28}
\and
A.~J.~Banday\inst{77, 7, 63}
\and
R.~B.~Barreiro\inst{52}
\and
J.~G.~Bartlett\inst{3, 54}
% ...
}
\institute{\small
Aalto University Mets\"{a}hovi Radio Observatory, Mets\"{a}hovintie 114, FIN-02540 Kylm\"{a}l\"{a}, Finland\\
\and
Agenzia Spaziale Italiana Science Data Center, c/o ESRIN, via Galileo Galilei, Frascati, Italy\\
\and
Astroparticule et Cosmologie, CNRS (UMR7164), Universit\'{e} Denis Diderot Paris 7, B\^{a}timent Condorcet, 10 rue A. Domon et L\'{e}onie Duquet, Paris, France\\
\and
Astrophysics Group, Cavendish Laboratory, University of Cambridge, J J Thomson Avenue, Cambridge CB3 0HE, U.K.\\
\and
Atacama Large Millimeter/submillimeter Array, ALMA Santiago Central Offices, Alonso de Cordova 3107, Vitacura, Casilla 763 0355, Santiago, Chile\\
\and
CITA, University of Toronto, 60 St. George St., Toronto, ON M5S 3H8, Canada\\
\and
CNRS, IRAP, 9 Av. colonel Roche, BP 44346, F-31028 Toulouse cedex 4, France\\
\and
California Institute of Technology, Pasadena, California, U.S.A.\\
\and
Centre of Mathematics for Applications, University of Oslo, Blindern, Oslo, Norway\\
\and
DAMTP, University of Cambridge, Centre for Mathematical Sciences, Wilberforce Road, Cambridge CB3 0WA, U.K.\\
\and
DSM/Irfu/SPP, CEA-Saclay, F-91191 Gif-sur-Yvette Cedex, France\\
\and
DTU Space, National Space Institute, Juliane Mariesvej 30, Copenhagen, Denmark\\
\and
Departamento de F\'{\i}sica, Universidad de Oviedo, Avda. Calvo Sotelo s/n, Oviedo, Spain\\
\and
Department of Astronomy and Astrophysics, University of Toronto, 50 Saint George Street, Toronto, Ontario, Canada\\
\and
Department of Physics \& Astronomy, University of British Columbia, 6224 Agricultural Road, Vancouver, British Columbia, Canada\\
\and
Department of Physics, Gustaf H\"{a}llstr\"{o}min katu 2a, University of Helsinki, Helsinki, Finland\\
\and
Department of Physics, Princeton University, Princeton, New Jersey, U.S.A.\\
\and
Department of Physics, Purdue University, 525 Northwestern Avenue, West Lafayette, Indiana, U.S.A.\\
\and
Department of Physics, University of California, Berkeley, California, U.S.A.\\
\and
Department of Physics, University of California, One Shields Avenue, Davis, California, U.S.A.\\
\and
Department of Physics, University of California, Santa Barbara, California, U.S.A.\\
\and
Department of Physics, University of Illinois at Urbana-Champaign, 1110 West Green Street, Urbana, Illinois, U.S.A.\\
\and
Dipartimento di Fisica G. Galilei, Universit\`{a} degli Studi di Padova, via Marzolo 8, 35131 Padova, Italy\\
\and
Dipartimento di Fisica, Universit\`{a} La Sapienza, P. le A. Moro 2, Roma, Italy\\
\and
Dipartimento di Fisica, Universit\`{a} degli Studi di Milano, Via Celoria, 16, Milano, Italy\\
\and
Dipartimento di Fisica, Universit\`{a} degli Studi di Trieste, via A. Valerio 2, Trieste, Italy\\
\and
Dipartimento di Fisica, Universit\`{a} di Ferrara, Via Saragat 1, 44122 Ferrara, Italy\\
\and
Dipartimento di Fisica, Universit\`{a} di Roma Tor Vergata, Via della Ricerca Scientifica, 1, Roma, Italy\\
\and
Discovery Center, Niels Bohr Institute, Blegdamsvej 17, Copenhagen, Denmark\\
\and
Dpto. Astrof\'{i}sica, Universidad de La Laguna (ULL), E-38206 La Laguna, Tenerife, Spain\\
\and
European Southern Observatory, ESO Vitacura, Alonso de Cordova 3107, Vitacura, Casilla 19001, Santiago, Chile\\
\and
European Space Agency, ESAC, Planck Science Office, Camino bajo del Castillo, s/n, Urbanizaci\'{o}n Villafranca del Castillo, Villanueva de la Ca\~{n}ada, Madrid, Spain\\
\and
European Space Agency, ESTEC, Keplerlaan 1, 2201 AZ Noordwijk, The Netherlands\\
\and
Haverford College Astronomy Department, 370 Lancaster Avenue, Haverford, Pennsylvania, U.S.A.\\
\and
Helsinki Institute of Physics, Gustaf H\"{a}llstr\"{o}min katu 2, University of Helsinki, Helsinki, Finland\\
\and
INAF - Osservatorio Astrofisico di Catania, Via S. Sofia 78, Catania, Italy\\
\and
INAF - Osservatorio Astronomico di Padova, Vicolo dell'Osservatorio 5, Padova, Italy\\
\and
INAF - Osservatorio Astronomico di Roma, via di Frascati 33, Monte Porzio Catone, Italy\\
\and
INAF - Osservatorio Astronomico di Trieste, Via G.B. Tiepolo 11, Trieste, Italy\\
\and
INAF/IASF Bologna, Via Gobetti 101, Bologna, Italy\\
\and
INAF/IASF Milano, Via E. Bassini 15, Milano, Italy\\
\and
INRIA, Laboratoire de Recherche en Informatique, Universit\'{e} Paris-Sud 11, B\^{a}timent 490, 91405 Orsay Cedex, France\\
\and
Imperial College London, Astrophysics group, Blackett Laboratory, Prince Consort Road, London, SW7 2AZ, U.K.\\
\and
Infrared Processing and Analysis Center, California Institute of Technology, Pasadena, CA 91125, U.S.A.\\
\and
Institut N\'{e}el, CNRS, Universit\'{e} Joseph Fourier Grenoble I, 25 rue des Martyrs, Grenoble, France\\
\and
Institut d'Astrophysique Spatiale, CNRS (UMR8617) Universit\'{e} Paris-Sud 11, B\^{a}timent 121, Orsay, France\\
\and
Institut d'Astrophysique de Paris, CNRS UMR7095, Universit\'{e} Pierre \& Marie Curie, 98 bis boulevard Arago, Paris, France\\
\and
Institute of Astronomy and Astrophysics, Academia Sinica, Taipei, Taiwan\\
\and
Institute of Astronomy, University of Cambridge, Madingley Road, Cambridge CB3 0HA, U.K.\\
\and
Institute of Theoretical Astrophysics, University of Oslo, Blindern, Oslo, Norway\\
\and
Instituto de Astrof\'{\i}sica de Canarias, C/V\'{\i}a L\'{a}ctea s/n, La Laguna, Tenerife, Spain\\
\and
Instituto de F\'{\i}sica de Cantabria (CSIC-Universidad de Cantabria), Avda. de los Castros s/n, Santander, Spain\\
\and
Istituto di Fisica del Plasma, CNR-ENEA-EURATOM Association, Via R. Cozzi 53, Milano, Italy\\
\and
Jet Propulsion Laboratory, California Institute of Technology, 4800 Oak Grove Drive, Pasadena, California, U.S.A.\\
\and
Jodrell Bank Centre for Astrophysics, Alan Turing Building, School of Physics and Astronomy, The University of Manchester, Oxford Road, Manchester, M13 9PL, U.K.\\
\and
Kavli Institute for Cosmology Cambridge, Madingley Road, Cambridge, CB3 0HA, U.K.\\
\and
LERMA, CNRS, Observatoire de Paris, 61 Avenue de l'Observatoire, Paris, France\\
\and
Laboratoire AIM, IRFU/Service d'Astrophysique - CEA/DSM - CNRS - Universit\'{e} Paris Diderot, B\^{a}t. 709, CEA-Saclay, F-91191 Gif-sur-Yvette Cedex, France\\
\and
Laboratoire Traitement et Communication de l'Information, CNRS (UMR 5141) and T\'{e}l\'{e}com ParisTech, 46 rue Barrault F-75634 Paris Cedex 13, France\\
\and
Laboratoire de Physique Subatomique et de Cosmologie, CNRS/IN2P3, Universit\'{e} Joseph Fourier Grenoble I, Institut National Polytechnique de Grenoble, 53 rue des Martyrs, 38026 Grenoble cedex, France\\
\and
Laboratoire de l'Acc\'{e}l\'{e}rateur Lin\'{e}aire, Universit\'{e} Paris-Sud 11, CNRS/IN2P3, Orsay, France\\
\and
Lawrence Berkeley National Laboratory, Berkeley, California, U.S.A.\\
\and
Max-Planck-Institut f\"{u}r Astrophysik, Karl-Schwarzschild-Str. 1, 85741 Garching, Germany\\
\and
Max-Planck-Institut f\"{u}r Radioastronomie, Auf dem H\"{u}gel 69, 53121 Bonn, Germany\\
\and
MilliLab, VTT Technical Research Centre of Finland, Tietotie 3, Espoo, Finland\\
\and
National University of Ireland, Department of Experimental Physics, Maynooth, Co. Kildare, Ireland\\
\and
Niels Bohr Institute, Blegdamsvej 17, Copenhagen, Denmark\\
\and
Observational Cosmology, Mail Stop 367-17, California Institute of Technology, Pasadena, CA, 91125, U.S.A.\\
\and
Optical Science Laboratory, University College London, Gower Street, London, U.K.\\
\and
SISSA, Astrophysics Sector, via Bonomea 265, 34136, Trieste, Italy\\
\and
SUPA, Institute for Astronomy, University of Edinburgh, Royal Observatory, Blackford Hill, Edinburgh EH9 3HJ, U.K.\\
\and
School of Physics and Astronomy, Cardiff University, Queens Buildings, The Parade, Cardiff, CF24 3AA, U.K.\\
\and
Space Sciences Laboratory, University of California, Berkeley, California, U.S.A.\\
\and
Spitzer Science Center, 1200 E. California Blvd., Pasadena, California, U.S.A.\\
\and
Stanford University, Dept of Physics, Varian Physics Bldg, 382 Via Pueblo Mall, Stanford, California, U.S.A.\\
\and
Universidad de Chile, Casilla 36-D, Santiago, Chile\\
\and
Universit\'{e} de Toulouse, UPS-OMP, IRAP, F-31028 Toulouse cedex 4, France\\
% ...
}

\abstract{Anomalous microwave emission (AME) has been observed by
  numerous experiments in the frequency range $\sim
  10$--$60$\,GHz. Using \Planck~maps and multi-frequency ancillary
  data, we have constructed spectra for two known AME regions: the
  Perseus and $\rho$~Ophiuchi molecular clouds. The spectra are well
  fitted by a combination of free-free radiation, cosmic microwave
  background, thermal dust, and electric dipole radiation from small
  spinning dust grains. The spinning dust spectra are the most
  precisely measured to date, and show the high frequency side clearly
  for the first time. The spectra have a peak in the range
  $20$--$40$~GHz and are detected at high significances of
  $17.1\sigma$ for Perseus and $8.4\sigma$ for $\rho$ Ophiuchi. In Perseus, spinning
  dust in the dense molecular gas can account for most of the AME; the
  low density atomic gas appears to play a minor role. In $\rho$ Ophiuchi, the
  $\sim 30$\,GHz peak is dominated by dense molecular gas, but there
  is an indication of an extended tail at frequencies
  $50$--$100$\,GHz, which can be accounted for by irradiated low
  density atomic gas. The dust parameters are consistent with those
  derived from other measurements.  We have also searched the
  \Planck~map at 28.5\,GHz for candidate AME regions, by subtracting a
  simple model of the synchrotron, free-free, and thermal dust. We present
   spectra for two of the candidates; S140 and S235 are bright \ion{H}{ii}~regions that show
  evidence for AME, and are well fitted by spinning dust models.}
%\date{Received date, Accepted date}


\authorrunning{Planck Collaboration}
\titlerunning{\textit{Planck} Early Results. XX. New light on anomalous emission from spinning dust grains}

\maketitle   

\allearlypapers   % ensures correct referencing of Planck papers
    
%%%%%%%%%%%%%%%%%%%%%%%%%%%%%%%%%%%%%%%%%%%%%%%%%%%%%%%%%%%%%%%%%%%%%%%%%%%%%%%%%

\section{Introduction}
\label{sec:introduction}

Anomalous microwave emission (AME) is an additional component of diffuse foreground emission that cannot be easily explained by synchrotron, free-free, or thermal dust emission. AME has been observed by numerous experiments over the frequency range $\sim$10--60\,GHz and is found to be very closely correlated with far infrared (FIR) emission associated with thermal emission from dust grains \citep{Kogut1996,Leitch1997,deOliveira-Costa1997,Banday2003,Lagache2003,deOliveira-Costa2004,Finkbeiner2004,Finkbeiner2004b,Davies2006,Dobler2008,Miville-Deschenes2008,Gold2010,Ysard2010b}. Electric dipole radiation from small rapidly spinning dust grains, or ``spinning dust,'' is thought to be emitted in the microwave region of the spectrum\footnote{\cite{Erickson1957} was the first to suggest the possibility of radio emission from spinning dust grains.}. Theoretical models predict a peaked spectrum, at a range of possible frequencies ($\sim$10--150\,GHz) depending on the properties of the dust grains and their environment \citep{Draine98b,Ali-Hamoud2009,Dobler2009,Ysard2010a}. The spectrum is very different from those of the traditional diffuse emission expected at these frequency ranges; for example, free-free, synchrotron, and thermal dust emission have power-law--like spectra at microwave frequencies. For these reasons, spinning dust emission has become the best explanation for the AME, although other physical mechanisms could still be contributing at some level, such as hot free-free emission \citep{Leitch1997}, hard synchrotron radiation \citep{bennett2003b}, or magneto-dipole emission \citep{Draine1999}.

The most direct evidence for spinning dust grains has come from dedicated observations of known dust clouds in the interstellar medium \citep{Finkbeiner2002,Casassus2006,Dickinson2009,Scaife2009,Dickinson2009,Dickinson2010}. Perhaps the best example is that of the Perseus molecular cloud. \cite{Watson2005} observed the region with the COSMOSOMAS telescope covering $11$--$17$\,GHz. By combining data from radio surveys, {\it WMAP}, and DIRBE, they were able to show very clearly a strong excess at frequencies $\sim$10--60\,GHz and a shape similar to spinning dust models. Another clear example is that of the photodissociation region (PDR) associated with the $\rho$~Ophiuchi molecular dust cloud. \cite{Casassus2008} observed the region centred at $(l,b)=(353\fdg0,+17\fdg0)$ at 31\,GHz with the Cosmic Background Imager (CBI) and found a close connection between the centimetre-wavelength emission and the dust emission. Further analysis of the spectrum confirmed that there was significant excess emission that was consistent with spinning dust models. AME has also been detected within a star-forming region in the galaxy NGC\,6946 \citep{Murphy2010}, which appears to be well-fitted by spinning dust models \citep{Scaife2010}.

\Planck\footnote{\Planck\ (\url{http://www.esa.int/Planck}) is a project of the European Space Agency (ESA) with instruments provided by two scientific consortia funded by ESA member states (in particular the lead countries France and Italy), with contributions from NASA (USA) and telescope reflectors provided by a collaboration between ESA and a scientific consortium led and funded by Denmark.} data now represent the next step in measuring and understanding spinning dust emission. The additional data at 28.5, 44.1, and 70.3\,GHz (where spinning dust is expected to emit strongly) and the high frequency data at $100$--$857$\,GHz allow an accurate model for the thermal emission to be removed. This is crucial in defining a precise spectrum for the spinning dust, particularly at the higher frequencies where the smallest dust grains dominate. Accurate spectra for spinning dust could potentially prove to be an important diagnostic tool for measuring the properties of the dust (e.g., the size distribution, density and average electric dipole moment) as well as the physical environment in the surrounding region (e.g., the interstellar radiation field).

In this paper, we study two known AME regions, the Perseus and $\rho$~Ophiuchi molecular cloud regions, and provide spectra over the frequency range $0.4$--$3000$\,GHz. We show, for the first time, an accurate residual spectrum after removal of the free-free, cosmic microwave background (CMB), and thermal dust components. The AME is clearly detected as a convex spectrum that is well-fitted by theoretical models of spinning dust emission. We investigate the parameters for the spinning dust to see whether they are reasonable, given the conditions in these environments. The \Planck~data, in combination with ancillary data, are also used to search for new regions of AME. We present two such regions and show that these newly discovered regions can also be fitted with a contribution from spinning dust grains. 

In this paper, Sects.\,\ref{sec:planck_data} and \ref{sec:ancillary_data} describe the \Planck\ and ancillary data that were used to measure the spectra of AME regions. Section~\ref{sec:results_perseus} presents the results for the Perseus molecular cloud and Sect.\,\ref{sec:results_roph} for the $\rho$ Ophiuchi cloud.  Section~\ref{sec:modelling} describes preliminary modelling of the AME in terms of spinning dust. Section~\ref{sec:new_regions} presents the search for new AME regions and gives two clear examples. Conclusions are given in Sect.\,\ref{sec:conclusions}.
 
\begin{table*}
\begin{center}
\caption{Summary of the data. \label{tab:data} }
\begin{tabular}{ccccl}
\hline
Frequency     &Telescope/    &Angular resolution      &Coverage    &Notes \\
$[$GHz$]$     &survey        &[arcmin]                &            &     \\
\hline
0.408         &JB/Eff/Parkes\tablefootmark{a} &$\approx 60$              &Full-sky   &NCSA desourced/destriped version \tablefootmark{b} \\
0.820         &Dwingeloo\tablefootmark{c}     &72              &Northern sky ($-7\degr<\delta<+85\degr$)  &Available on the web\tablefootmark{d} \\
1.420          &Stockert/Villa-Elisa\tablefootmark{e}    &36    &Full-sky &Courtesy of W. Reich  \\
2.326         &HartRAO\tablefootmark{f}        &20             &Southern sky ($-83\degr<\delta<+13/32^{\circ}$)&Courtesy of J. Jonas (priv. comm.)  \\
$11$--$17$        &COSMOSOMAS\tablefootmark{g}     &$\approx 60$     &Northern sky ($+24\degr<\delta<+44\degr$)  &Filtered on large angular scales \\
22.8          &{\it WMAP} 7-yr\tablefootmark{h}      &$\approx 49$   &Full-sky                         &$1^{\circ}$ smoothed version\tablefootmark{b} \\
28.5          &\Planck~LFI\tablefootmark{i}    &\getsymbol{LFI:FWHM:30GHz:units}   &Full-sky     & \\
33.0          &{\it WMAP} 7-yr\tablefootmark{h}      &$\approx 40$   &Full-sky                         &$1^{\circ}$ smoothed version\tablefootmark{b} \\ 
40.7          &{\it WMAP} 7-yr\tablefootmark{h}      &$\approx 31$   &Full-sky                         &$1^{\circ}$ smoothed version\tablefootmark{b} \\
44.1          &\Planck~LFI\tablefootmark{i}    &\getsymbol{LFI:FWHM:44GHz:units}   &Full-sky     & \\
60.7          &{\it WMAP} 7-yr\tablefootmark{h}      &$\approx 21$   &Full-sky                         &$1^{\circ}$ smoothed version\tablefootmark{b}  \\
70.3          &\Planck~LFI\tablefootmark{i}    &\getsymbol{LFI:FWHM:70GHz:units}   &Full-sky     & \\
93.5          &{\it WMAP} 7-yr\tablefootmark{h}      &$\approx 13$   &Full-sky                         &$1^{\circ}$ smoothed version\tablefootmark{b}  \\
100          &\Planck~HFI\tablefootmark{j}    &\getsymbol{HFI:FWHM:Mars:100GHz:units}   &Full-sky     & \\
143          &\Planck~HFI\tablefootmark{j}    &\getsymbol{HFI:FWHM:Mars:143GHz:units}   &Full-sky     & \\
217          &\Planck~HFI\tablefootmark{j}    &\getsymbol{HFI:FWHM:Mars:217GHz:units}   &Full-sky     & \\
353          &\Planck~HFI\tablefootmark{j}    &\getsymbol{HFI:FWHM:Mars:353GHz:units}   &Full-sky     & \\
545          &\Planck~HFI\tablefootmark{j}    &\getsymbol{HFI:FWHM:Mars:545GHz:units}   &Full-sky     & \\
857          &\Planck~HFI\tablefootmark{j}    &\getsymbol{HFI:FWHM:Mars:857GHz:units}   &Full-sky     & \\
1249          &{\it COBE}-DIRBE\tablefootmark{k}     &$\approx 40$   &Full-sky                         &LAMBDA website\tablefootmark{b}     \\
2141          &{\it COBE}-DIRBE\tablefootmark{k}     &$\approx 40$   &Full-sky                         &LAMBDA website\tablefootmark{b}      \\
2997          &{\it COBE}-DIRBE\tablefootmark{k}     &$\approx 40$   &Full-sky                         &LAMBDA website\tablefootmark{b}      \\
\hline
\end{tabular}
\tablefoot{
\tablefoottext{a}{\cite{Haslam1982}}~\tablefoottext{b}{\url{http://lambda.gsfc.nasa.gov/}}~\tablefoottext{c}{\cite{Berkhuijsen1972}}~\tablefoottext{d}{\url{http://www.mpifr-bonn.mpg.de/survey.html}}~\tablefoottext{e}{\cite{Reich1982,Reich1986,Reich2001}}~\tablefoottext{f}{\cite{Jonas1998}}~\tablefoottext{g}{\cite{Watson2005}}~\tablefoottext{h}{\cite{Jarosik2010}}~\tablefoottext{i}{\cite{planck2011-1.6}}~\tablefoottext{j}{\cite{planck2011-1.7}}~\tablefoottext{k}{\cite{Hauser1998}}
}
\end{center}
\end{table*}
\endPlancktable
%\tablenote example \par
%}
%\endtable

%%%%%%%%%%%%%%%%%%%%%%%%%%%%%%%%%%%%%%%%%%%%%%%%%%%%%%%%%%%%%%%%%%%%%%%%%%%%%%%%%

 \section{\textit{Planck} data}
\label{sec:planck_data}

\Planck\ \citep{tauber2010a, planck2011-1.1} is the third-generation space mission to measure the anisotropy of the cosmic microwave background (CMB).  It observes the sky in nine frequency bands covering 30--857\,GHz with high sensitivity and angular resolution from 31\arcm\ to 5\arcm.  The Low Frequency Instrument (LFI; \citealt{Mandolesi2010, Bersanelli2010, planck2011-1.4}) covers the 30, 44, and 70\,GHz bands with amplifiers cooled to 20\,\hbox{K}.  The High Frequency Instrument (HFI; \citealt{Lamarre2010, planck2011-1.5}) covers the 100, 143, 217, 353, 545, and 857\,GHz bands with bolometers cooled to 0.1\,\hbox{K}.  Polarisation is measured in all but the highest two bands \citep{Leahy2010, Rosset2010}.  A combination of radiative cooling and three mechanical coolers produces the temperatures needed for the detectors and optics \citep{planck2011-1.3}.  Two data processing centres (DPCs) check and calibrate the data and make maps of the sky \citep{planck2011-1.7, planck2011-1.6}.  \Planck's sensitivity, angular resolution, and frequency coverage make it a powerful instrument for Galactic and extragalactic astrophysics as well as cosmology.  Early astrophysics results are given in Planck Collaboration, 2011h--z.

In this paper, we start with the \Planck~(non CMB-subtracted) maps at nominal frequencies of 28.5, 40.1, 70.3, 100, 143, 217, 353, 545, and 857\,GHz (see Table~\ref{tab:data}). For the spectral analyses, we smoothed the HEALPix maps to a common angular resolution of $1\degr$ and degraded to $N_{\rm side}=512$ \citep{Gorski2005}. We assume Gaussian beams and use the average beamwidths given in \cite{planck2011-1.7,planck2011-1.6}. Details of the beam, such as the exact shape and variation of the beam across the sky, do not significantly affect the results presented here, due to the applied smoothing which dominates the spatial response of the maps. Furthermore, extended emission is less sensitive to the details of the beam. We convert from CMB thermodynamic units to Rayleigh-Jeans brightness temperature units using the standard conversion factors given in \cite{planck2011-1.7} and \cite{planck2011-1.6}. Colour corrections to account for the finite bandpass at each frequency are applied during the modelling of the spectra for each object; these are typically a few per cent for LFI and $10$--$15\,\%$ for HFI. The 100\,GHz data are significantly contaminated by the CO $J\!=\!1\!\rightarrow\!0$ line at 115\,GHz. We apply a nominal correction to the 100\,GHz map using the CO survey of \cite{Dame2001}. We multiply the integrated $1\degr$-smoothed line intensity map (in units of K~km~s$^{-1}$) by $14.2$ \citep{planck2011-1.7} and subtract it from the 100~GHz map. Instead of attempting to correct for other lines that may contaminate the 217 and 353\,GHz data, we do not include these channels in the modelling of spectra if they are observed to be in excess by more than $2\sigma$, relative to the best-fit model when omitting these frequencies. Uncertainties are taken to be the nominal uncertainties in absolute calibration and gain (given in the aforementioned papers), which are added in quadrature. If these are below $3\,\%$, we conservatively assume an overall uncertainty of $3\,\%$. 

Jack-knife tests, in which the data were split into two halves (in terms of time and also HFI detectors; see \citealt{planck2011-1.7}), provided a check of the consistency of the \Planck~data. The tests confirmed that the results were consistent with each other to within a small fraction of the derived uncertainties.

%%%%%%%%%%%%%%%%%%%%%%%%%%%%%%%%%%%%%%%%%%%%%%%%%%%%%%%%%%%%%%%%%%%%%%%%%%%%%%%%%

\section{Ancillary data}
\label{sec:ancillary_data}

Although \Planck~observes over a wide frequency range ($30$--$857$\,GHz), it is still important to include ancillary data to extend the frequency range into the FIR, and in particular, to lower frequencies. To supplement \Planck~data we therefore use a range of publicly available surveys and data available within the team. Table~\ref{tab:data} summarises the data used in the analysis and the origin of the maps that were used. In the following sections, we describe the various data sets in more detail.

\subsection{Low frequency radio data}

The radio data consist of a number of low frequency surveys in the range $0.408$--$2.326$\,GHz (Table~\ref{tab:data}).  We used the most up-to-date versions of the data sets, and where necessary, regridded the maps into the HEALPix format \citep{Gorski2005} by using a procedure which computes the surface intersection between individual pixels of the survey with the intersecting HEALPix pixels. This procedure has been shown to conserve photometry (Paradis et al., in prep.). We note that there are significant baseline uncertainties (i.e., offsets) in these maps, but these will not affect our results since we are subtracting a local background for each object, which removes any such offset. The data include the full-sky map at 408~MHz \citep{Haslam1982}, the 820~MHz Dwingeloo survey \citep{Berkhuijsen1972}, the new full-sky map at 1420~MHz \citep{Reich1982,Reich1986,Reich2001}, and the 2326~MHz HartRAO southern survey \citep{Jonas1998}. The 1420~MHz map was multiplied by a factor of 1.55 to account for the main beam to full beam ratio \citep{Reich1988}, and to bring the calibration in line with the other data. We assume a $10\,\%$ overall calibration uncertainty for these data.

\subsection{COSMOSOMAS data}

The COSMOSOMAS experiments \citep{Gallegos2001} consisted of two circular scanning instruments COSMO11 \citep{Hildebrandt2007} and COSMO15 \citep{Fernandez2006} which operated at the Teide observatory (altitude 2400\,m, Tenerife). They have produced $0\fdg8$--$1\fdg1$ resolution maps of $\approx$10,000~deg$^2$ in four frequency bands (10.9, 12.7, 14.7, 16.3\,GHz) that are ideal for filling the gap between low frequency radio surveys and the lowest frequency channels of {\it WMAP} and \Planck. The average sensitivity of each map is in the range $80$--$120\,\upmu$K per beam, and the observing strategy is described in \cite{Gallegos2001}. As COSMOSOMAS was a ground-based experiment, there is significant contamination from atmospheric and receiver $1/f$ noise, which needs to be filtered out. This results in a loss of response to angular scales larger than 5\degr\ and introduces low level correlations over the scan ring. This must be taken into account when considering diffuse emission at angular scales $\gtrsim 1\degr$ and this will be described in Sect.\,\ref{sec:results_perseus}. A $10\,\%$ overall calibration uncertainty is assumed.

\subsection{WMAP data}

{\it WMAP} 7-year data are included in the analysis \citep{Jarosik2010}. The data span $23$--$94$\,GHz and thus complement \Planck\ data, particularly the K-band (22.8\,GHz) channel. We use the $1\degr$-smoothed maps available from the LAMBDA website. We apply colour corrections to the central frequencies using the recipe described by \cite{Jarosik2003} and assume a conservative $3\,\%$ overall calibration uncertainty.

\subsection{DIRBE data}

To sample the peak of the blackbody curve for temperatures $\gtrsim 15$\,K we include the {\it COBE}-DIRBE data at $240\,\upmu$m (1249\,GHz), $140\,\upmu$m (2141\,GHz) and $100\,\upmu$m (2997\,GHz). The DIRBE data are the Zodi-Subtracted Mission Average (ZSMA) maps \citep{Hauser1998} regridded into the HEALPix format. Colour corrections were applied as described in the DIRBE explanatory supplement version 2.3. Data at higher frequencies are not considered, since these will be dominated by transiently heated grains which are not in thermal equilibrium with the interstellar radiation field and therefore cannot be modelled with a single modified blackbody curve. Furthermore, at wavelengths $\lesssim40\,\upmu$m the spectrum is dominated by many emission/absorption lines.

%The {\it IRAS} data are the re-processed versions, IRIS, of \cite{Miville-Deschenes2005}. Colour corrections are applied as described in the DIRBE and IRAS explanatory supplements.

\begin{figure*}
\centering
\includegraphics[width=1.0\textwidth]{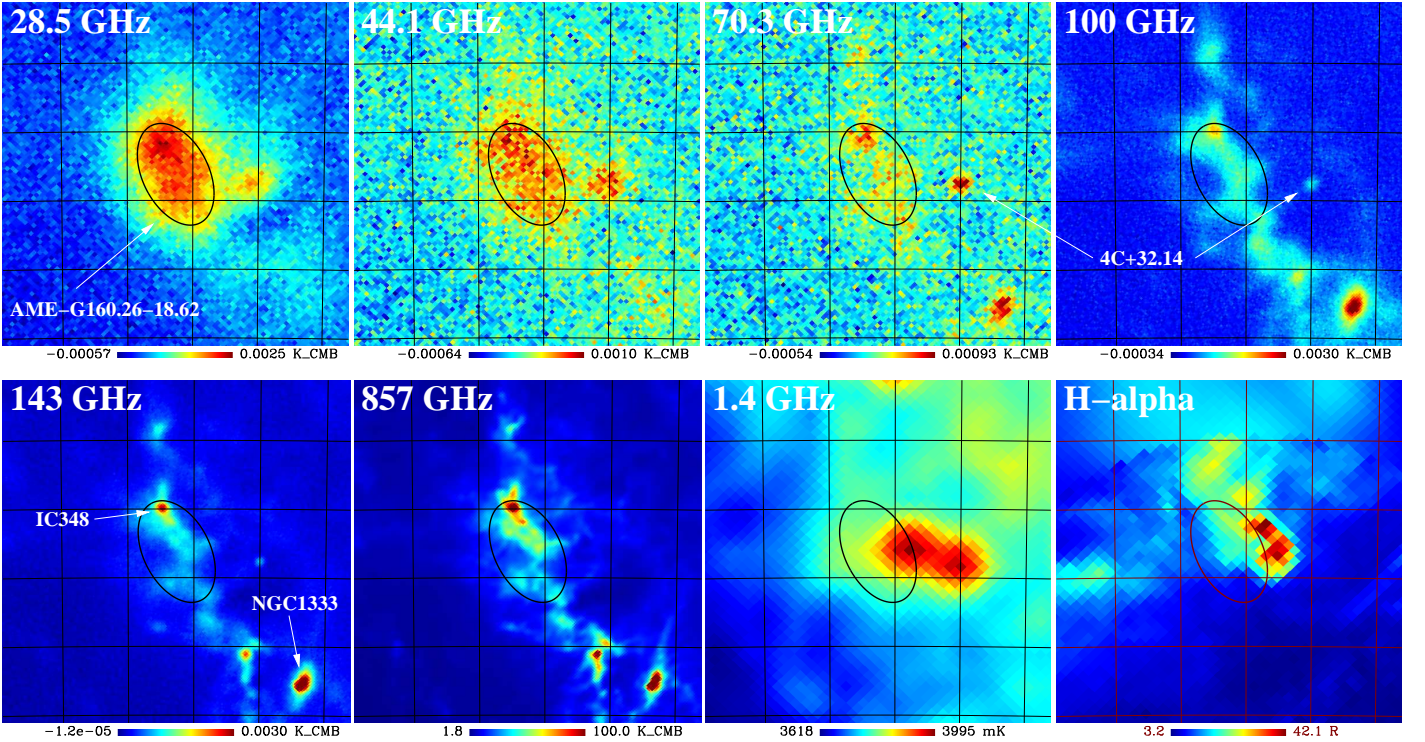}
\caption{Maps of the Perseus molecular cloud region at their original angular resolution. From left to right, top row: \planck~28.5; 44.1; 70.3 and 100\,GHz. Bottom row: \planck~143 and 857\,GHz; 1.4\,GHz; and H$\alpha$. The maps cover $5\degr \times 5\degr$ centred on $(l,b)=(160.\!\degr26,-18.\!\degr62)$ and have linear colour scales. The graticule has $1\degr$ spacing in Galactic coordinates. The FWHM of the elliptical Gaussian model used to fit the flux density in the filtered maps (see text) is shown. The strong AME is evident at $30$--$70$\,GHz. \label{fig:perseus_planck_maps}}
\end{figure*}

\begin{figure*}
\centering
\includegraphics[width=1.0\textwidth]{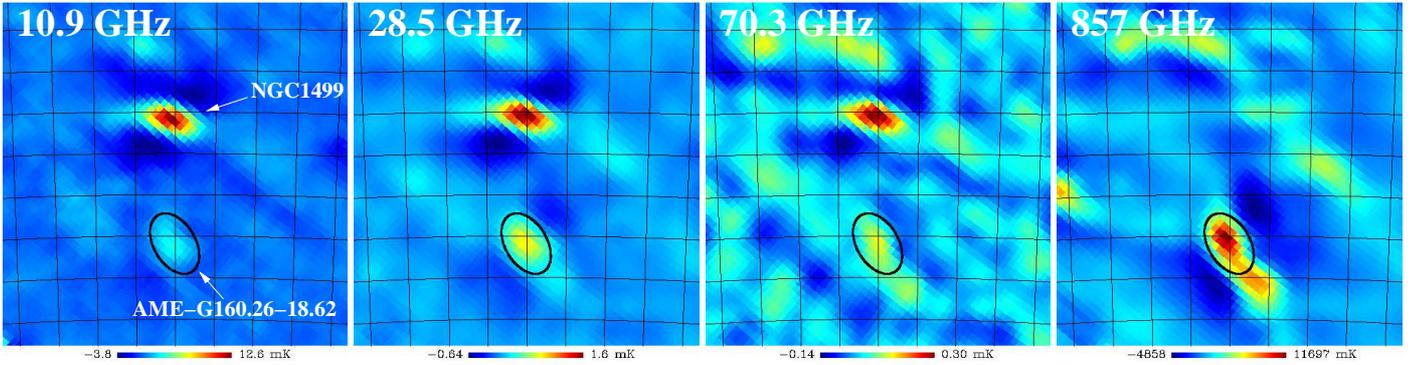}
\caption{$16\degr \times 16\degr$ maps after filtering with the COSMOSOMAS data reduction pipeline. The maps are centred at $(l,b)=(160\degr,-15\degr)$ and the graticule spacing is $2\degr$ in Galactic coordinates. From left to right: COSMOSOMAS 10.9\,GHz; \planck~28.5\,GHz; 70.3\,GHz; and 857\,GHz. The elliptical Gaussian model used to calculate the spectrum is indicated. \label{fig:filtered_maps} } 
\end{figure*}

%%%%%%%%%%%%%%%%%%%%%%%%%%%%%%%%%%%%%%%%%%%%%%%%%%%%%%%%%%%%%%%%%%%%%%%%%%%%%%%%%

\section{Perseus Molecular Cloud}
\label{sec:results_perseus}

\subsection{Introduction and Maps}

The Perseus molecular cloud complex is a relatively nearby giant molecular cloud of $1.3\times10^4$\,M$_{\odot} $ at a distance of 260\,pc \citep{Cernicharo1985} having an angular extent of $6\degr\times 2\degr$. It contains a chain of six dense cores (B\,5, IC\,348, B\,1, NGC\,1333, L\,1455 and L\,1448) with $A_v>2$\,mag running northeast--southwest and with two main centres of medium-mass star formation associated with the reflection nebulae IC\,348 and NGC\,1333 near each end (see Fig.~\ref{fig:perseus_planck_maps}). As a nearby example of low-to-medium mass star formation it has been extensively studied, with the {\it Spitzer} c2d Legacy Program \citep{Evans2003} and the COMPLETE survey \citep{Ridge2003}, which provide a wealth of infrared and spectral line data to understand conditions throughout the cloud.

Anomalous emission was first detected as an excess in this region of the sky via the Tenerife experiment \citep{Davies1987} at 5\degr~angular scales, and  it contributed most of the dust-correlated signal found by \cite{deOliveira1999}. With the 1\degr~beam and $11$--$17$\,GHz frequency bands of the COSMOSOMAS experiment \citep{Gallegos2001} it was possible to locate and identify the Perseus molecular cloud as the strong source of this emission \citep{Watson2005}. Follow-up observations with the VSA interferometer \citep{Tibbs2010} at 10\arcmin~resolution detected five ``hot-spots'' of anomalous emission, one of which corresponds to the dense region ($A_v \gtrsim 10$) around the IC~348 core. The other four hot-spots are associated with nearby dense dust knots that lie on a warm ring of dust emission (G159.6$-$18.5), about $90$\arcmin~to the south-east of IC~348. This ring, which is seen at $100\,\upmu$m,  is thought to be warmed by the OB star HD\,278942 \citep{deZeeuw1999}, which is also responsible for a weak \ion{H}{ii} region in the centre of the ring, and which can be seen in the low-frequency continuum surveys with an integrated flux density of a few janskys. The AME appears to have little or no polarisation, with an upper limit of $1.8\,\%$ at 33\ GHz \citep{Lopez-Caraballo2010}.

\Planck~maps of the Perseus molecular cloud region, covering 30--857\,GHz, are shown in Fig.~\ref{fig:perseus_planck_maps}, along with the 1.4\,GHz and H$\alpha$ maps. For display purposes the \Planck~maps have been CMB-subtracted as described in \cite{planck2011-1.6} and \cite{planck2011-1.7} and the units are thermodynamic kelvins. The strong dust-correlated AME at $30$--$70$\,GHz is evident; it has no obvious counterpart at 1.4\,GHz but correlates well with the higher ($>$100\,GHz) \planck~frequencies, which are dominated by thermal dust. The majority of the AME comes from the northeast end of the molecular cloud, which harbours the dense IC~348 reflection nebula. Some weak free-free emission is seen to the east and the contribution from the extragalactic radio source 4C+32.14 is clear and must be removed (Sect.\,\ref{sec:perseus_spectrum}). The H$\alpha$ (and 1.4\,GHz emission) morphology is not co-located with the emission seen at $30$--$70$\,GHz, but is likely to be significantly absorbed by the intervening dust. Nevertheless, the brightest H$\alpha$ pixels predict $\approx 0.25$\,mK at 30\,GHz \citep{Dickinson2003}, a factor of ten below the AME at 30\,GHz. It is interesting that there is considerable bright structure observed in the \planck~HFI maps, which is not always associated with strong AME. This is likely to be a consequence of a different environment and dust grain population.

\subsection{Spectrum}
\label{sec:perseus_spectrum}

To reduce correlated $1/f$ noise from the atmosphere/receiver the COSMOSOMAS time-ordered data have been filtered with the suppression of the first seven harmonics in the FFT of the circular scans. To compare with flux densities from other instruments one must either correct for the flux lost in filtering or filter all the other data in the same way. With extended sources such as Perseus there is uncertainty in estimating a single correction factor to take into account the filtering process, which in turn directly affects the shape of the frequency spectrum. Filtering all the data in the same way retains the spectral shape and ensures that the spatial frequencies used to form the spectrum are common between the different data sets.

The first step is to smooth all the data to the same resolution as the lowest frequency channel of the COSMOSOMAS data, i.e., a FWHM of $1\fdg12$. Each data set is then filtered using the COSMOSOMAS reduction pipeline and binned into a map as was done with the original COSMOSOMAS data \citep{Fernandez2006,Hildebrandt2007}. Example maps at 10.9, 28.5, 70.3, and 857\,GHz are shown in Fig.~\ref{fig:filtered_maps}. To estimate the flux density at each frequency the filtered maps were compared to a filtered model  map, which was passed through the same scan strategy. For the model we use an elliptical Gaussian ($1\fdg6\times1\fdg0$ with major-axis PA $51\degr$), centred at $(\textrm{RA}, \textrm{Dec}) = (55\fdg33,+31\fdg67)$ or $(l,b)=(160\fdg26,-18\fdg62)$, following \cite{Watson2005}; this gives a solid angle $\Omega=5.52 \times 10^{-4}$\,sr. We will refer to this as AME-G160.26$-$18.62. There is a flat-spectrum radio quasar, 4C+32.14 (NRAO\,140), within 1\degr, with a flux density of $\approx 1$--$3$\,Jy, which is seen in the \Planck~Early Release Compact Source Catalogue (ERCSC) at frequencies up to 143\,GHz \citep{planck2011-1.10}. We use the flux densities and beam widths taken from the ERCSC and subtract the two-dimensional Gaussian profile from the maps before smoothing and filtering. Meanwhile in the COSMOSOMAS maps we must extrapolate the flux density of 4C+32.14 to the appropriate frequencies and simulate its contribution to the filtered map; we assume $2.0$\,Jy at $11$--$19$\,GHz, which is in good agreement with NVSS \citep{Condon1998} and recent 2.7\,GHz measurements made with the Effelsberg telescope \citep{Reich2009}. The contribution of 4C+32.14 is $\approx 5\,\%$ of the integrated flux density,  so the final results are not very sensitive to the exact treatment of the subtraction or the variability of the source. Both the smoothing and filtering result in correlated noise between the pixels in the map. A correlation analysis \citep[e.g.,][]{Davies2006} is used to calculate the flux density taking into account the correlated noise. The covariance matrix is estimated from the size of the smoothing kernel and the point source response to the scan strategy. A simulated map is produced for the source, with an integrated flux density of 1\,Jy, and a maximum likelihood method is used to find the amplitude and error for the given frequency channel data within a $19 \times 19$ pixel box ($6\fdg 3 \times 6\fdg 3$) around the source position.

\begin{figure}
\centering
\includegraphics[width=0.35\textwidth,angle=-90]{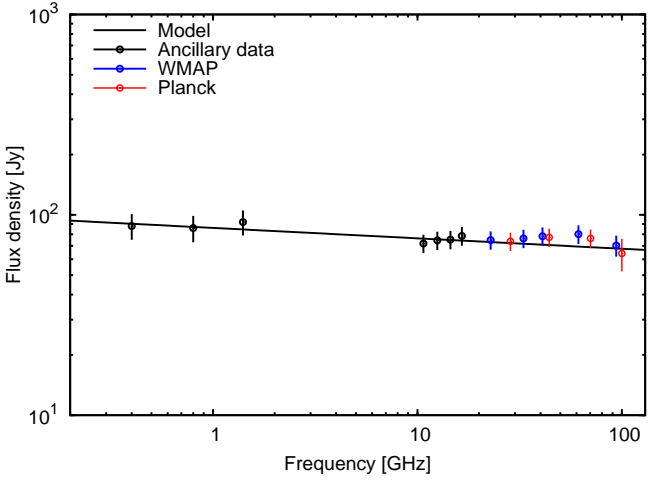}
\caption{Spectrum of the California nebula (NGC1499), measured using the filtered flux method (see text). A simple power-law fit is shown. \label{fig:california}}
\end{figure}

As a control, we tested the simulation and filtering pipeline by applying the technique to the extended California nebula \ion{H}{ii} region (NGC\,1499), approximately $6\degr$ to the north of Perseus at $(l,b)=(160\fdg 60,-12\fdg 05)$, as seen in Fig.~\ref{fig:filtered_maps}. The emission at radio/microwave wavelengths is dominated by optically thin free-free emission. Flux densities at frequencies $>$100\,GHz were affected by dust emission lying adjacent to the north-west of NGC\,1499 and thus were not included. The spectrum is shown in Fig.~\ref{fig:california}. A simple power-law fit, with a best-fitting spectral index $\alpha=-0.05\pm0.01$, is overplotted. The fit is relatively good ($\chi^2$/dof$=1.23$) but there is some evidence for AME at frequencies $\approx 30$--$100$\,GHz. Furthermore, the spectral index is slightly flatter than the value expected for optically thin free-free emission over this frequency range, $\alpha \approx -0.1$ to $-0.15$ \citep{Dickinson2003}. Indeed, AME has been observed to be significant in some \ion{H}{ii} regions \citep{Dickinson2009,Todorovic2010}. We confirmed the spectral shape with a standard aperture photometry analysis (as described in Sect.\,\ref{sec:results_roph}) and will investigate this in more detail in a future publication. Note that the excess at $30$--$100$\,GHz disappears when the spectrum is calculated away from the nearby dust feature to the north-west of NGC\,1499. Given the consistency of the flux densities between the frequencies, and that we have obtained similar results using standard aperture photometry, we are confident that the spectrum calculated using the filtered maps is robust.

\begin{figure}
\centering
\includegraphics[width=0.35\textwidth,angle=-90]{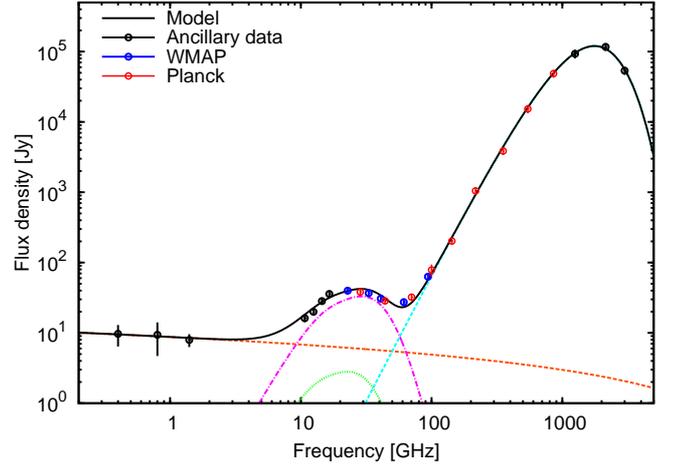}
\caption{Spectrum of AME-G160.26$-$18.62 in the Perseus molecular cloud. The best-fitting model consisting of free-free (orange dashed line), spinning dust, and thermal dust (light blue dashed line) is shown. The two-component spinning dust model consists of high density molecular gas (magenta dot-dashed line) and low density atomic gas (green dotted line). \label{fig:perseus_sed} } 
\end{figure}

The spectrum for AME-G160.26$-$18.62 in Perseus is shown in Fig.~\ref{fig:perseus_sed}. The flux densities and associated errors are listed in Table~\ref{tab:perseus}. The spectrum is well sampled across the radio, microwave and FIR regimes and is a significant improvement on that presented in \cite{Watson2005}, with the additional \planck~data allowing a much more accurate spectrum of the AME to be extracted. The low-frequency ($<2$\,GHz) data show a flat spectrum consistent with free-free emission while the high frequencies ($>$100\,GHz) are dominated by thermal dust emission. The excess at $\approx 10$--$70$\,GHz is evident and has a peaked (convex) spectrum, with a maximum at 25\,GHz. 

\begin{figure}
\centering
\includegraphics[width=0.35\textwidth,angle=-90]{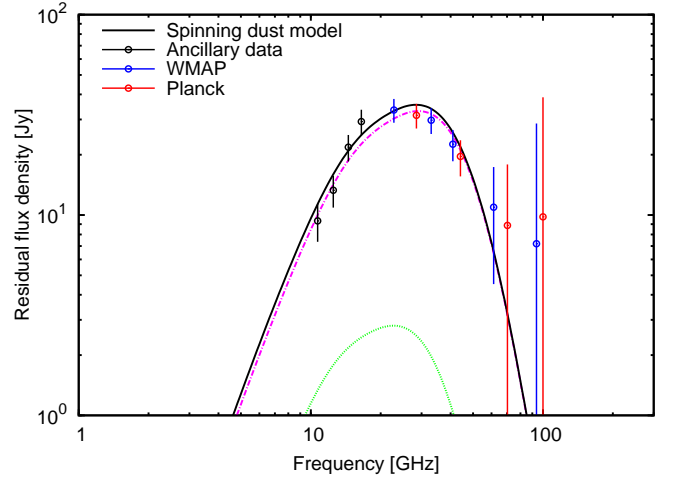}
\caption{Spectrum of AME-G160.26$-$18.62 in the Perseus molecular cloud after subtracting the best-fit free-free, CMB and thermal dust components. A theoretical spinning dust model, consisting of two components is shown as the black solid line; the magenta dot-dashed line is for high density molecular gas and the green dotted line is for low density atomic gas, which is a very small contribution in this case (see Sect.\,\ref{sec:modelling}). \label{fig:perseus_residual}}
\end{figure}

\begin{table}
\begin{center}
\caption{Flux densities for AME-G160.26$-$18.62 in the Perseus molecular cloud and residual fluxes when free-free, CMB and thermal dust components are removed. The fitted spinning dust model consists of two components.}
\begin{tabular}{ccc}
\hline
Frequency   &Flux density     &Flux density residual \\
$[$GHz$]$   &[Jy]                &[Jy]   \\
\hline
 0.408 &$9.7 \pm 3.3$ &$0.2 \pm 3.4 $  \\
 0.82 &$9.4 \pm 4.7$ &$0.5 \pm 4.8 $  \\
 1.42 &$8.0 \pm 1.7$ &$-0.5 \pm 1.8 $  \\
 10.9 &$16.1 \pm 1.8$ &$9.4 \pm 2.0 $  \\
 12.7 &$20.0 \pm 2.2$ &$13.3 \pm 2.4 $  \\
 14.7 &$28.4 \pm 3.1$ &$21.8 \pm 3.3 $  \\
 16.3 &$35.8 \pm 4.0$ &$29.3 \pm 4.2 $  \\
 22.8 &$39.8 \pm 4.2$ &$33.4 \pm 4.5 $  \\
 28.5 &$38.1 \pm 4.0$ &$31.5 \pm 4.4 $  \\
 33.0 &$36.7 \pm 3.9$ &$29.7 \pm 4.4 $  \\
 40.9 &$30.8 \pm 3.3$ &$22.6 \pm 4.0 $  \\
 44.1 &$28.5 \pm 3.2$ &$19.6 \pm 4.0 $  \\
 61.3 &$27.4 \pm 3.4$ &$10.9 \pm 6.4 $  \\
 70.3 &$32.2 \pm 3.9$ &$8.9 \pm 9.0 $  \\
 93.8 &$63.0 \pm 7.8$ &$7 \pm 21 $  \\
 100 &$78 \pm 15$ &$10 \pm 29 $  \\
 143 &$202 \pm 22$ &$-25 \pm 80 $  \\
 217 &$1050 \pm 130$ &$120 \pm 320 $  \\
 353 &$3860 \pm 470$ &$-600 \pm 1400 $  \\
 545 &$15300 \pm 2100$ &$-800 \pm 4900 $  \\
 857 &$48700 \pm 6100$ &$-1000 \pm 14000 $  \\
 1250 &$93000 \pm 13000$ &$-2400 \pm 28000 $  \\
 2143 &$117000 \pm 15000$ &$7000 \pm 35000 $  \\
 2997 &$53600 \pm 6700$ &$-2000 \pm 20000 $  \\
\hline
\end{tabular}
\label{tab:perseus}
\end{center}
\end{table}

The data allow us to fit a multi-component parametric model to the flux density spectrum. The model consists of four components: free-free emission; thermal dust emission; a CMB fluctuation; and spinning dust emission. The sum is
\begin{equation}
S = S_{\rm ff} + S_{\rm td} + S_{\rm cmb} + S_{\rm sp}.
\end{equation}
The free-free flux density, $S_{\rm ff}$, is calculated from the brightness temperature, $T_{\rm ff}$, based on the optical depth, $\tau_{\rm ff}$, using the standard formulae:
\begin{equation}
S_{\rm ff} = \frac{2  k T_{\rm ff}  \Omega  \nu^2}{c^2},
\end{equation} 
where $k$ is the Boltzmann constant, $\Omega$ is the solid angle, and $\nu$ is the frequency,
\begin{equation}
T_{\rm ff} = T_{e}(1-e^{-\tau_{\rm ff}}),
\end{equation}
and the optical depth, $\tau_{\rm ff}$, is given by
\begin{equation}
\tau_{\rm ff} = 3.014 \times 10^{-2}  T_{e}^{-1.5}  \nu^{-2}   {\rm EM}  g_{\rm ff},
\end{equation}
where $T_{e}$ is the electron temperature (in units of K), EM is the emission measure (in units of cm$^{-6}$~pc) and $g_{\rm ff}$ is the Gaunt factor, which is approximated as
\begin{equation}
g_{\rm ff} = \ln \left(\frac{4.955 \times 10^{-2}}{\nu/{\rm GHz}}\right) + 1.5 \, \ln(T_{e}).
\end{equation}
It is the Gaunt factor that results in a slight steepening of the free-free spectrum with frequency, which is particularly pronounced at $\gtrsim 100$\,GHz. The thermal dust is modelled as a single-component, modified blackbody curve, $\nu^{\,\beta}B(\nu,T_{\rm d})$, which we normalise using an optical depth at $250\,\upmu$m (1.2~THz), $\tau_{250}$, as
\begin{equation}
S_{\rm td} =  2 \, h \, \frac{\nu^3}{c^2} \frac{1}{e^{h\nu/kT_{\rm d}}-1} \, \tau_{250} \, (\nu/1.2\,\textrm{THz})^{\,\beta} \,\Omega,
\end{equation}
where $h$ is the Planck constant and $\beta$ is the emissivity index. The CMB anisotropy is evaluated as the differential of the blackbody temperature function, in thermodynamic units ($\Delta T_{\rm CMB}$), converted to flux density units,  
\begin{equation}
S_{\rm cmb} = \left(\frac{2  k \Omega \nu^2}{c^2}\right) \Delta T_{\rm CMB}.
\end{equation}
The spinning dust model is based on an assumed theoretical emissivity curve, $j_{\nu}$ (in units of Jy\,sr$^{-1}$\,cm$^2$\,H$^{-1}$), normalised by the average column density per H nucleon, $N_{\rm H}$, and solid angle, $\Omega$,
\begin{equation}
S_{\rm sp} = N_{\rm H} \, j_{\nu} \,\Omega .
\end{equation}
As described in Sect.\,\ref{sec:modelling}, we use the SPDUST code \citep{Ali-Hamoud2009} to model the spinning dust spectrum.

The best-fitting model is plotted in Fig.~\ref{fig:perseus_sed}. The fitted parameters are EM for free-free, differential temperature of CMB perturbation ($\Delta T_{\rm CMB}$), dust temperature ($T_{\rm d}$), dust emissivity index ($\beta$) and $250~{\upmu}$m  opacity ($\tau_{250}$). The spinning dust model consists of two components, to represent dust associated with dense molecular gas and low density atomic gas, and will be discussed in Sect.\,\ref{sec:modelling}. The electron temperature is assumed to have a typical value of 8000\,K for gas in the solar neighbourhood. A conservative $3\,\%$ uncertainty in the CO correction is assumed at 100\,GHz and is not included in the fit. Similarly, the 217\,GHz value is not included in the fit, due to the possible small contamination from the CO $J\!=\!2\!\rightarrow\!1$ line. From the initial spectral fit the spectral index is found at each frequency, which is then used to calculate the appropriate colour corrections for each band in an iterative manner. After applying the colour corrections (typically a few per cent correction), the spectrum is fitted again to obtain the final flux densities, which are given in Table~\ref{tab:perseus}. The uncertainty in the fitted model for the spectrum is computed by propagating the errors of each of the parameters and combining them with the other sources of error, i.e., background/noise residuals and an overall calibration error.  The fit is very good with $\chi^2/\textrm{dof}=0.95$. The best-fitting parameters are: $\textrm{EM}=148\pm13.5$~cm$^{-6}$\,pc for free-free; $T_{\rm d}=18.5\pm0.6$~K and $\beta=1.65 \pm 0.08$ for the thermal dust; and a negligible CMB contribution of $\Delta T_{\rm CMB}=-6\pm66\,\upmu$K.

In Fig.~\ref{fig:perseus_residual} we show the residual spectrum after subtracting the free-free, CMB and thermal dust models.  The additional uncertainty from the subtraction of the models is added in quadrature to the flux density errors to find the final uncertainties on the residual flux densities, which are listed in Table~\ref{tab:perseus}. This represents the most precise spectrum of AME to date. The weighted average over the range 10 to 94\,GHz gives a detection significance level of $17.1\sigma$. The peak of the spinning dust component is centred at $\approx 25$\,GHz and the low and high frequency sides are well-defined in the range $10$--$90$\,GHz. A physical model for the spinning dust is presented in Sect.\,\ref{sec:modelling}. It consists of two components (atomic and molecular), which are overplotted in Fig.~\ref{fig:perseus_residual}, and is seen to be an excellent fit to the data.

%%%%%%%%%%%%%%%%%%%%%%%%%%%%%%%%%%%%%%%%%%%%%%%%%%%%%%%%%%%%%%%%%%%%%%%%%%%%%%%%%

\section{$\rho$ Ophiuchi Molecular Cloud}
\label{sec:results_roph}

\subsection{Introduction and Maps}

The $\rho$~Ophiuchi molecular cloud \citep[e.g.,][]{Encrenaz1974,Kulesa2005, Young2006} lies in the Gould Belt of
the closest molecular complexes, at a distance $D = 135
\pm$15\,pc \citep{Perryman1997}.  It is
undergoing intermediate-mass star formation. Ultra-violet radiation from its
hottest young stars heats and dissociates exposed layers, but does not
ionise bulk hydrogen. The most prominent PDR in $\rho$~Ophiuchi is the $\rho$~Oph~W filament, with $n_{\rm H} = 10^4$--$10^5$~cm$^{-3}$, which has been
studied by \cite{Liseau1999} and \cite{Habart2003}. $\rho$~Oph~W is excited by HD~147889
(a B2,3~IV binary), the earliest star in the
$\rho$~Oph star formation complex. The line of sight to HD~147889 can
be taken to characterise the physical conditions in the ambient
diffuse cloud, with $N_{\rm H} \sim 2 \times 10^{20}$~cm$^{-2}$, $n_{\rm H}
\sim 400~$~cm$^{-3}$; no accurate values are available for
  $N_{\rm H_2}$ \citep{vanDishoeck1989,Kazmierczak2010}. The
bulk of the mass in the $\rho$~Ophiuchi cloud is situated in the Oph~A
molecular core \citep[e.g.,][]{Wootten1978,Wootten1979}, located roughly 10\arcmin\ southwest of $\rho$~Oph~W.

Only faint radiation from the Rayleigh-Jeans tail of $\sim 10$--$100$~K
dust is expected at wavelengths longwards of $\sim$3\,mm. Yet CBI
observations revealed that the $\rho$~Oph~W PDR is surprisingly bright
at centimetre wavelengths \citep{Casassus2008}. The {\em WMAP} 33\,GHz image confirms that the general
location of the signal seen by CBI is offset by $10\arcmin$ from the
centroid of the {\em WMAP}~94\,GHz data, located on the Oph~A core. The
signal seen at centimetre wavelengths has no 5\,GHz counterpart;
instead the optically thin free-free emission is located in the
immediate vicinity of HD\,147889. The other early-type stars in the
complex are S\,1 (B4~V) and SR\,3 (B6~V), and each is surrounded by bright IR emission peaks, which  lack any radio
counterpart.

\begin{figure*}
\centering
\includegraphics[width=1.0\textwidth]{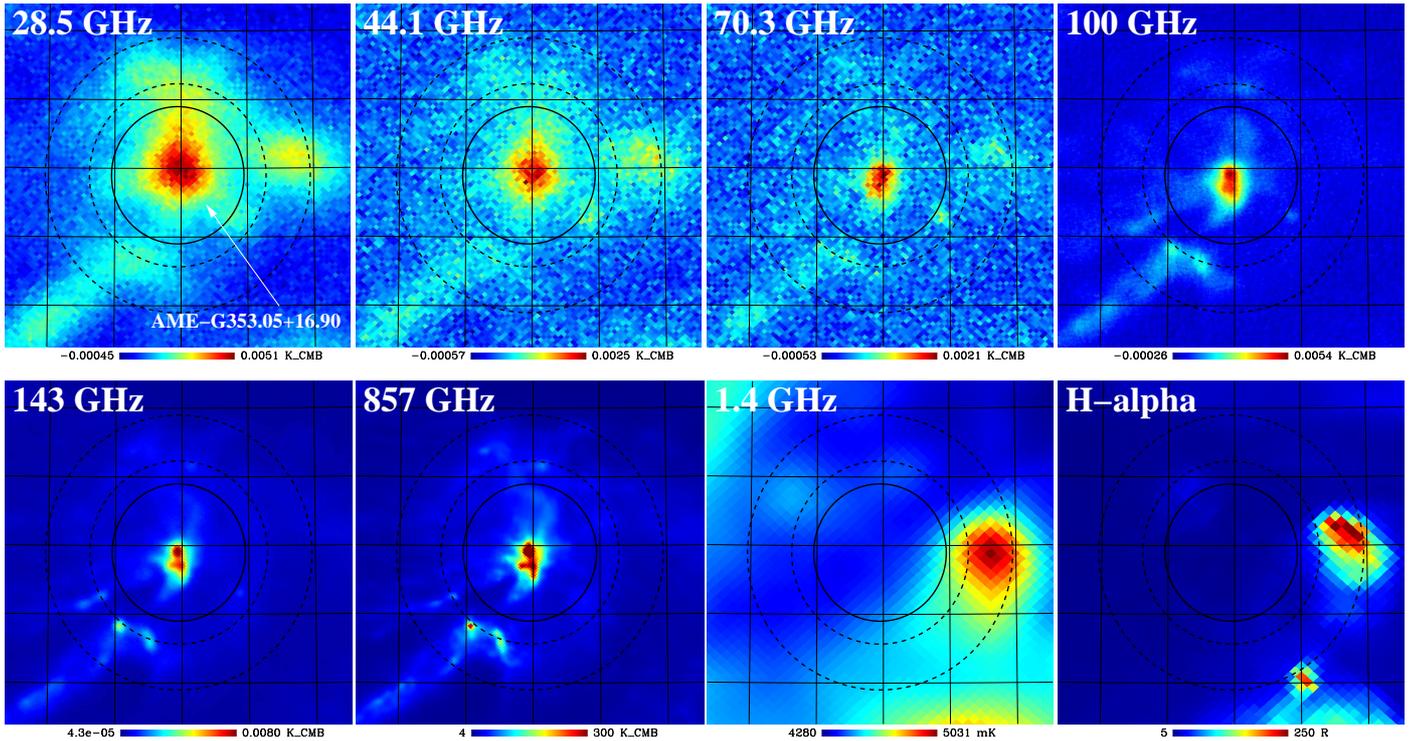}
\caption{Maps of the $\rho$ Ophiuchi molecular cloud region at their original angular resolution. From left to right, top row: \planck~28.5; 44.1; 70.3; and 100\,GHz. Bottom row: \planck~143 and 857\,GHz; 1.4\,GHz; and H$\alpha$. The maps cover $5\degr \times 5\degr$ centred on $(l,b)=(353\fdg05,+16\fdg90)$ and have linear colour scales. The graticule has $1\degr$ spacing in Galactic coordinates. The circular aperture and background annulus, which were used to calculate the flux density (see text), are indicated. The strong AME is evident at 28.5 and 44.1\,GHz. \label{fig:roph_planck_maps}}
\end{figure*}

\Planck~maps of the $\rho$ Ophiuchi molecular cloud region covering 30--857\,GHz, along with the 1.4\,GHz and H$\alpha$ maps, are shown in Fig.~\ref{fig:roph_planck_maps}. Strong AME is evident  at 28.5--44.1\,GHz, with the bulk of the emission falling within $0.\!\degr5$ of the central position. For display purposes the \Planck~maps have been CMB-subtracted as described in \cite{planck2011-1.6} and \cite{planck2011-1.7} and the units are thermodynamic kelvins. The \ion{H}{ii} region to the west is also clearly detected in the \Planck~LFI maps, but is well separated from the bulk of the AME. In the location of the strongest AME, there is a valley in intensity seen at low frequencies, supported by little H$\alpha$ emission. The AME is well correlated with \Planck~HFI maps, indicating an association with dust. 

\begin{figure}
\begin{center}
\includegraphics[width=0.5\textwidth]{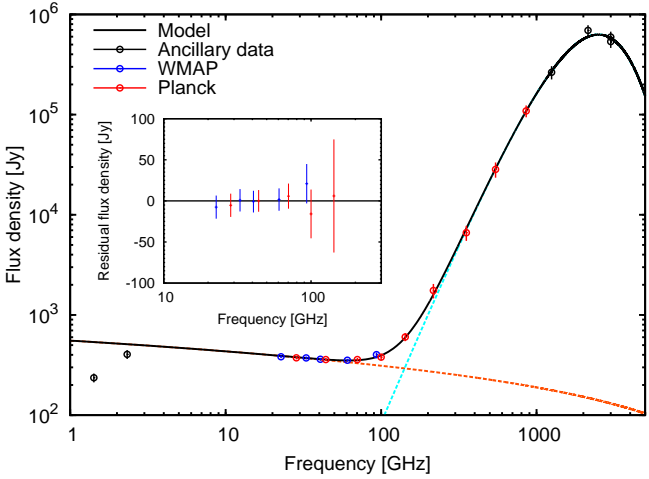}
\caption{Spectrum of the Orion nebula (M42) \ion{H}{ii} region using aperture photometry. The model consists of optically thin free-free emission (orange dashed line) and thermal dust emission (light blue dashed line). The inner panel shows the residual spectrum after removal of the model, indicating the consistency between \Planck~and {\it WMAP} data. \label{fig:M42_spec}}
\end{center}
\end{figure}

\subsection{Spectrum}
\label{sec:roph_spectrum}

The $\rho$ Ophiuchi molecular cloud (at declination $-24\degr$) lies outside the range of the COSMOSOMAS survey. We therefore chose to use simple aperture photometry on the unfiltered maps to calculate the spectrum. The maps were first smoothed to a common resolution of $1\degr$ and converted to units of Jy~pixel$^{-1}$. The brightness in each pixel was then summed over a given aperture to give the flux density. We chose a circular aperture centred at $(l,b)=(353\fdg 05,16\fdg90)$ with a radius of $60\arcmin$; this corresponds to a solid angle $\Omega=9.57\times10^{4}$\,sr. We will refer to this source as AME-G353.05+16.90. An estimate of the background brightness is subtracted using the median value in a circular annulus between an inner radius of $80\arcmin$ and an outer radius of $120\arcmin$, as indicated in Fig.~\ref{fig:roph_planck_maps}. This removes local background emission and any residual offsets in the maps. Errors are estimated from the standard deviation in the background annulus, which was verified with simulations of random sources of known flux density that were injected into the sky maps. The results were not strongly sensitive to the exact choice of background annulus; all values were within the $1\sigma$ uncertainties. This  indicates that the emission in the aperture is bright relative to the local background fluctuations. We also compared two halves of the data as a jack-knife test. Results were found to be consistent to within a fraction of the derived uncertainties. 

We tested the aperture photometry by measuring the spectrum of well-known, bright objects. We chose bright \ion{H}{ii} regions, which are expected to be dominated by thermal bremsstrahlung (free-free) radiation at radio frequencies and thermal dust in the sub-mm/FIR. Fig.~\ref{fig:M42_spec} shows the spectrum of the Orion nebula (M42) with an aperture of radius 60\arcmin~and background annulus between radii of $80\arcmin$ and $120\arcmin$.  At frequencies below a few gigahertz, the free-free emission is becoming optically thick, with a turnover frequency of $\approx 2$\,GHz. At frequencies $\sim$10--60\,GHz, the spectrum is well-fitted by a simple optically thin free-free spectrum ($\alpha \approx -0.1$). The {\it WMAP} and \Planck~data points are in good agreement to within $\approx 1\,\%$; the residuals are consistent with zero and within a fraction of the assumed uncertainties. At frequencies $\gtrsim 100$\,GHz, thermal dust emission dominates and is well described by a modified blackbody with an emissivity index $\beta=1.71 \pm 0.12$ and dust temperature $T_{\rm d}=25.7 \pm 1.7$~K. 

We note that the lack of AME in M42 is likely to be due to the difference in conditions (e.g., intense radiation field) that may result in depletion of the smallest dust grains, which would result in a large reduction in the amount of spinning dust emission. This is a different situation to the more diffuse warm ionised medium (WIM) studied by \cite{Dobler2009}, where they detected AME in the WIM at the level of $\approx 20\,\%$ relative to the free-free emission.

\begin{figure}
\begin{center}
\includegraphics[width=0.35\textwidth,angle=-90]{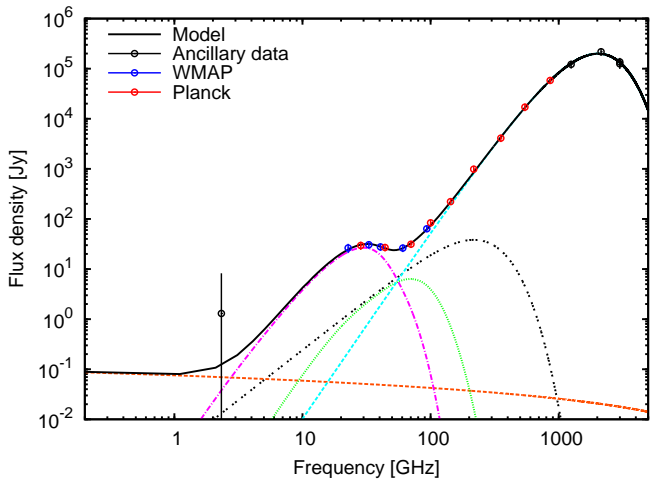}
\caption{Spectrum of AME-G353.05+16.90 in the $\rho$ Ophiuchi West molecular cloud. The best-fitting model consisting of free-free (orange dashed line), spinning dust, CMB (black double/triple-dotted line), and thermal dust (light blue dashed line), is shown. The spinning dust model consists of two components: high density molecular gas (magenta dot-dashed line); and low density atomic gas (green dotted line). The 100/217\,GHz data are contaminated by CO line emission and are not included in the fit. \label{fig:roph_sed}}
\end{center}
\end{figure}

Fig.~\ref{fig:roph_sed} shows the spectrum of AME-G353.03+16.90 in the $\rho$ Oph West region; the flux densities are listed in Table~\ref{tab:roph}. The same model used for the Perseus region (Sect.\,\ref{sec:perseus_spectrum}) was fitted to the data, except for the details of the spinning dust component, which will be discussed in Sect.\,\ref{sec:modelling}. Significant CO line contamination is visible at 100 and 217\,GHz, so these two bands were not included in the fit. There is minimal free-free emission within the aperture, and hence the error at low ($0.4$--$2.3$\,GHz) frequencies is dominated by fluctuations in the local background, as indicated by the large error in the 2.3\,GHz flux density. Note that the 408/1420~MHz data were included in the fit but were consistent with zero. The best-fitting parameters are: ${\rm EM}=1\pm48~$cm$^{-6}~$pc; $T_{\rm d}=20.7\pm1.6$~K; $\beta=1.75\pm0.18$; and CMB contribution $\Delta T_{\rm CMB}=82 \pm 61\,\upmu$K. The fit of the overall model is very good ($\chi^2/{\rm dof}=0.29$).

As a test of the impact that the \Planck~data was having on the spectrum, we repeated the analysis omitting the \Planck\ data. As expected, the impact was considerable. In particular, the thermal dust model is constrained much more strongly the HFI data are included. For example, the uncertainty in the dust temperature increases to $\pm 2.5$\,K, while the uncertainty in the emissivity index increases to $\pm 0.36$. This is crucial for making precise measurements of the AME at frequencies, particularly above $50$\,GHz where the subtraction of thermal dust is critical.

\begin{figure}
\centering
\includegraphics[width=0.35\textwidth,angle=-90]{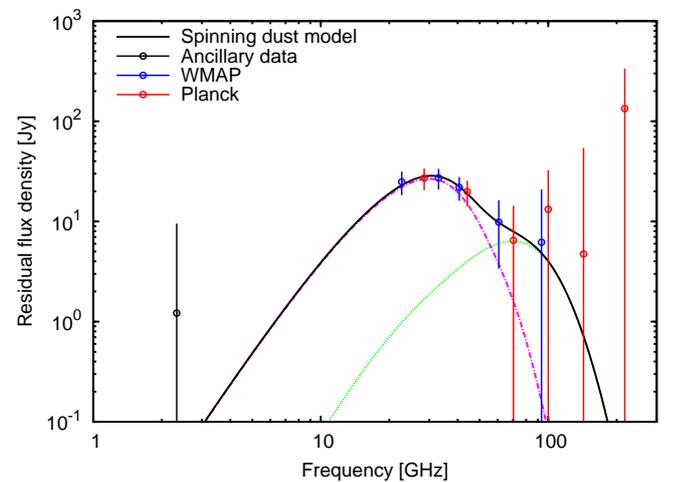}
\caption{Spectrum of AME-G353.05+16.90 in the $\rho$ Ophiuchi molecular cloud after subtracting the best-fit free-free, CMB and thermal dust components. A theoretical spinning dust model consisting of two components is shown as the black solid line; the magenta dot-dashed line is for dense molecular gas and the green dotted line is for irradiated low density atomic gas (see Sect.\,\ref{sec:modelling}). The 100/217\,GHz data are contaminated by CO line emission and are not included in the fit. \label{fig:roph_residual}}
\end{figure}

\begin{table}
\begin{center}
\caption{Flux densities for AME-G353.05+16.90 in the $\rho$ Ophiuchi molecular cloud and residual flux densities when free-free, CMB and thermal dust components are removed.}
\begin{tabular}{ccc}
\hline
Frequency   &Flux density     &Flux density residual \\
$[$GHz$]$   &[Jy]                &[Jy]   \\
\hline
 0.4 &$-6.8 \pm 9.4$ &$-7 \pm 11 $  \\
 1.4 &$-1.0 \pm 6.8$ &$-1.0 \pm 8.3 $  \\
 2.3 &$1.3 \pm 6.9$ &$1.2 \pm 8.3 $  \\
 22.7 &$26.3 \pm 5.5$ &$24.8 \pm 6.6 $  \\
 28.5 &$29.6 \pm 5.6$ &$27.3 \pm 6.6 $  \\
 33.0 &$30.7 \pm 5.3$ &$27.2 \pm 6.3 $  \\
 40.7 &$27.7 \pm 4.6$ &$21.9 \pm 5.8 $  \\
 44.1 &$27.0 \pm 4.4$ &$19.9 \pm 5.7 $  \\
 60.6 &$26.3 \pm 4.5$ &$9.8 \pm 6.5 $  \\
 70.3 &$31.4 \pm 5.1$ &$6.5 \pm 7.8 $  \\
 93.4 &$63.6 \pm 8.9$ &$6 \pm 15 $  \\
 100 &$84 \pm 13$ &$13 \pm 19 $  \\
 143 &$222 \pm 31$ &$5 \pm 49 $  \\
 217 &$989 \pm 16$ &$130 \pm 200 $  \\
 353 &$4100 \pm 670$ &$-180 \pm 780 $  \\
 545 &$17100 \pm 2900$ &$400 \pm 3200 $  \\
 857 &$58300 \pm 8800$ &$500 \pm 10000 $  \\
 1249 &$122000 \pm 22000$ &$-5000 \pm 25000 $  \\
 2141 &$218000 \pm 37000$ &$20000 \pm 41000 $  \\
 2997 &$125000 \pm 26000$ &$-9000 \pm 32000 $  \\
    \hline
\end{tabular}
\label{tab:roph}
\end{center}
\end{table}

In Fig.~\ref{fig:roph_residual} we show the residual spectrum after subtracting the free-free, CMB, and thermal dust models.  The additional error from the subtraction of the models is added in quadrature to the flux density errors to obtain the uncertainty on the residual flux densities, listed in Table~\ref{tab:roph}. As in Perseus, the residual spectrum has a clearly defined convex shape, peaking at a frequency of 30\,GHz; the excess is significant at the $8.4\sigma$ level. The spectrum at $50$--$100$\,GHz is flatter than a single spinning dust model can easily account for, suggesting a different environment or distribution of dust grains. This may also be due to variations in the dust emissivity index due to multiple dust components (e.g., \citealt{Finkbeiner1999}). To test this possibility would require the inclusion of the 100/217~GHz data after careful subtraction of the CO line contamination.

A theoretical spinning dust model consisting of two components (associated with atomic and molecular gas) is overplotted and provides an excellent fit to the data. Note that the denser molecular gas represents the dominant AME with peak at $\approx 30$\,GHz, while the irradiated low density atomic gas accounts for the $50$--$100$\,GHz part of the spectrum. The physical model is presented in Sect.\,\ref{sec:modelling}.

%%%%%%%%%%%%%%%%%%%%%%%%%%%%%%%%%%%%%%%%%%%%%%%%%%%%%%%%%%%%%%%%%%%%%%%%%%%%%%%%%

% TABLE 1 SpDust parameters 
%------------------------------------------------------------------------------------------
\begin{table}[t]
  \centering
  \caption{Parameters in the spinning dust models for Perseus and $\rho$ Ophiuchi: $x_{\rm H}$ and $x_{\rm C}$
    are the abundances of H$^+$ and C$^+$; $y$ is the molecular
    fraction $2n_{\rm H_2}/n_{\rm H}$; $a_0$ is the centroid of the PAH size distribution; $b_{\rm C}$ is the PAH abundance of
    carbon in PAHs; and $z$ is the depth of the emission region along the
    line of sight. The radiation
    field is that of \cite{Mathis1983} multiplied by the factor $G_0$. The ionised gas
    accounts for the free-free emission but does not contribute to the
    spinning dust emission.   \label{tab:model_param} }
  \begin{tabular}{l c c c }
   \hline           
   \hline           
   Gas state & Molecular & Atomic &  Ionised \\
   \hline
 %\multicolumn{4}{c}{Perseus}     \\
%Perseus  & & &  \\ 
          & &Perseus  &  \\
\hline
   $N_{\rm H}$ [10$^{21}$~cm$^{-2}$]     & 11.7  & 1.3   &  0.4\\
   $n_{\rm H}$ [cm$^{-3}$]              & 250   & 30     &  1 \\
   $z$ [pc]                            & 15.1  & 14.0  & \ldots \\
   $G_0$                               & 1     & 2     & \ldots \\
   $T$ [K]                            & 40    & 100    & $8 \times 10^3$\\
   $x_{\rm H}$ [ppm]                   & 112   & 410    &  10$^6$\\
   $x_{\rm C}$ [ppm]                   & $<$1  & 100    & \ldots  \\
   $y$                                & 1     &  0.1   & \ldots  \\
   $a_0$ [nm]                         & 0.58  & 0.53   & \ldots  \\
   $b_{\rm C}$ [ppm]                   & 68    &68      & \ldots \\
   $\beta$                            & \ldots&1.65    & \ldots \\ 
   $T_{\rm d}$ [K]                     & \ldots&18.5    & \ldots \\
   $\tau_{250}$                        & \ldots&$9.4 \times 10^{-4}$ & \ldots \\
   \hline
 %\multicolumn{4}{c}{$\rho$ Ophiuchi} \\
%$\rho$ Ophiuchi    & & &  \\ 
          &   &$\rho$ Ophiuchi   &  \\ 
\hline
   $N_{\rm H}$ [10$^{21}$~cm$^{-2}$]    & 18.2   & 0.4  & 0.4 \\
   $n_{\rm H}$ [cm$^{-3}$]             & $2 \times 10^4$ & 200 &  0.5\\
   $z$ [pc]                          & 0.3     & 0.6   & \ldots \\
   $G_0$                             & 0.4     & 400   & \ldots  \\
   $T$ [K]                           & 20      & 10$^3$ & $8 \times 10^3$\\
   $x_{\rm H}$ [ppm]                  & 9.2     & 373    & 10$^6$ \\
   $x_{\rm C}$ [ppm]                  & $<$1    & 100    & \ldots  \\
   $y$                               &1        &  0.1   & \ldots \\
   $a_0$ [nm]                        & 0.60    & 0.38   & \ldots  \\
   $b_{\rm C}$ [ppm]                  & 65      &50      & \ldots   \\
   $\beta$                           & \ldots  &1.75    & \ldots \\ 
   $T_{\rm d}$ [K]                    & \ldots  &20.7    & \dots  \\
   $\tau_{250}$                       & \ldots  &$3.2 \times 10^{-3}$ &  \ldots \\
   \hline
   \hline
  \end{tabular}
 \end{table}
%------------------------------------------------------------------------------------------

\section{Modelling the anomalous microwave emission with spinning dust}
\label{sec:modelling}

Comparisons of the AME with dust emission in localised regions
\citep{Casassus2006,Scaife2009,Tibbs2010,Castellanos2011} and in all-sky surveys \citep{Lagache2003,Ysard2010b}
suggest small dust grains (polycyclic aromatic hydrocarbons, PAH) as
carriers of this emission, supporting the spinning dust
explanation \citep{Draine98b}. The present and previous
measurements of the AME show that this component 
peaks at 20--40 GHz with a brightness of 0.01 to 0.1 MJy~sr$^{-1}$ (see
references in Sect.\,\ref{sec:introduction}). Current spinning dust models \citep{Ali-Hamoud2009,Ysard2010a,Hoang2010,Silsbee2011} indicate that low-density ($n_{\rm H}\leq$ 30 cm$^{-3}$), atomic
diffuse gas peaks in this frequency range. Therefore the AME has often been explained with spinning
dust emission arising mostly from regions of low density gas, such as from the Cold Neutral Medium (CNM) or Warm Neutral Medium (WNM). The column density
$N_{\rm H}$ associated with the AME can be derived from observations provided the emissivity of spinning dust is known:
adopting an average of 10$^{-23}$~MJy~sr$^{-1}$~cm$^2$~H$^{-1}$ from theoretical models, we find that $N_{\rm H}\negmedspace\sim\negmedspace10^{22}$\,cm$^{-2}$ for a flux density level of 0.1\,MJy\,sr$^{-1}$. For a gas density of 30\,cm$^{-3}$, the extent of the emitting region
is thus $N_{\rm H}/n_{\rm H}\sim$100 pc, much larger than the observed size of the present AME clouds ($< 10$~pc). The situation is even worse for the warm ionised medium (WIM) component ($n_{\rm H}<1$\,cm$^{-3}$). The spinning dust emission of these regions must therefore include a significant
contribution from gas which is denser by at least a factor of 10, as suggested by \cite{Casassus2008} and \cite{Castellanos2011}. 
In the preliminary models of the Perseus and $\rho$ Ophiuchi
regions presented below, we show that, assuming
plausible physical conditions and PAH size distributions,
spinning dust emission from dense gas can explain most of 
the AME. 

The rotation of PAHs in space is governed by IR emission
and gas-grain interactions \citep{Draine98b}.
Quantitative modelling of spinning dust emission requires the knowledge
of many ($\sim$10) parameters describing the state
of the gas, the radiation field, and the PAH
size distribution, as well as the electric dipole moment. 
As shown by \cite{Ali-Hamoud2009} and \cite{Ysard2010b}, the peak frequency of spinning dust emission in neutral gas
(and for  radiation intensity $G_0<10^2$, where $G_0$ is the multiplying factor relative to the mean radiation intensity in the interstellar medium, as defined by \citealt{Mathis1983}) depends chiefly on the gas
density and on the size of PAHs.\footnote{We assume here the electric dipole moment to be as in
  \citealt{Draine98b}, a prescription also shown to be compatible with
 the AME extracted from {\it WMAP} data \citep{Ysard2010b}.} The largest PAHs have the lowest emissivity. In low
temperature gas ($G_0 < 3$   and $n_{\rm H}>30$ cm$^{-3}$), 
the gas-grain collisions dominate the excitation and the damping of the grain rotation. Thus 
the spinning dust spectrum
becomes sensitive to the abundance of the major ions, H$^+$ and C$^+$, denoted $x_{\rm H}$
and $x_{\rm C}$ respectively. 
To compute the spinning dust emissivity, we use {\tt SPDUST}\footnote{\url{http://www.tapir.caltech.edu/~yacine/spdust/spdust.html}} in the
case 2 of \cite{Silsbee2011}. Rather than fitting all the spinning dust
parameters on the observed spectrum, we derive realistic values from observations, and fit the amplitude to the spectrum.
We derive $x_{\rm H}$ from the ionisation balance of hydrogen 
assuming a standard cosmic-ray ionisation rate of $5 \times 10^{-17}$~s$^{-1}$~H$^{-1}$ \citep{Williams1998,Wolfire2003}.
Conversely, $x_{\rm C}$ cannot be estimated so readily because
C undergoes reactions with H$_2$ and PAHs \citep{Roellig2006,Wolfire2008}; we therefore take $x_{\rm C}$ as
a free parameter. The size distribution of PAHs is
assumed to be a log-normal of centroid $a_0$ and width $\sigma=0.4$. The abundance of PAHs is set by $b_{\rm C}$, the
abundance of carbon locked up in PAHs.

Now we describe the physical conditions within our two regions, the Perseus and $\rho$ Ophiuchi molecular clouds, 
as derived from observations, and how we take them into account in our modelling  (see Table~\ref{tab:model_param}). For both regions, the free-free contribution is that of the \ion{H}{ii} region
and the thermal dust emission is described as a modified blackbody,
$I_{\nu}=\tau_{250}\,\left(\lambda/250\upmu{\rm
    m}\right)^{\,-\beta}\,B_{\nu}(T)$  where $\tau_{250}$ is the dust
opacity at 250\,$\upmu$m, $T$ the dust temperature (see Table~\ref{tab:model_param}), and
$B_\nu$ is the blackbody brightness. We do not include the contribution
of the ionised gas to the spinning dust emission because of its low
density and also because PAHs have been shown to be destroyed efficiently in such environments
\citep{Lebouteiller2007,Lebouteiller2011}. In addition, we note that the current {\tt SPDUST} code does not
  include ionising photons ($h\nu\geq 13.6$~eV), which
are important for the PAH charge and IR emission.

\subsection{Perseus molecular cloud}

At the degree scale, the Perseus line of sight intersects a molecular
cloud surrounding the \ion{H}{ii} region associated with the HD\,278942 B0\,V star.
\cite{Ridge2006} argued that this ionised shell has a density of $\sim 1$\,cm$^{-3}$.
The gas density and temperature of the molecular component are taken
from the analysis of observed C$_2$ lines carried out by \cite{Iglesias-Groth2011}\footnote{A single high resolution measurement may not be representative of the line-of-sight on degree scales, since the density and temperature may vary considerably across the beam.} 
The parameters of our model, which provides a good
match to the AME, are given in Table~\ref{tab:model_param}, and the
corresponding emission is shown in Fig.~\ref{fig:perseus_residual}. We note that the depth, $z$, of the
emissive region along the sightline is compatible with determinations
at smaller scales \citep{Ridge2006}. The
spinning dust emission is dominated by the contribution of the dense
gas; the low-density atomic gas appears to play only a minor role. This is consistent with the results of \cite{planck2011-7.3}, which show
that, on large scales, there is bright AME associated with molecular clouds traced by CO everywhere 
in the Galaxy, while the association with the diffuse and atomic phase
correlated with \ion{H}{i} is less clear. 

In contrast to previous work where $x_{\rm H}=0$ was assumed in molecular
gas, our ionisation balance of H gives $x_{\rm H}\sim 3\times 10^{-5}$, a
change particularly important for Perseus. We take $x_{\rm C}$ to be
1 ppm: at this level the C$^+$ ion has no influence on the spinning dust emission.
Adopting larger sizes for the  PAHs ($a_0=0.58$\,nm) than those used  in previous spinning dust models [but comparable to
the diffuse interstellar medium ($0.64$\,nm); \citealt{Compiegne2011}], the molecular gas accounts
well for the AME. 

\subsection{The $\rho$ Ophiuchi molecular cloud}

The $\rho$~Ophiuchi spectrum includes contributions from a
photo\-dissociated interface region (PDR) and the ionised gas associated with the star HD\,147889. 
We note that IR observations have shown that the PAH emission arises from dense irradiated 
interfaces \citep{Abergel2002,Habart2005}. 
As constrained by \cite{Habart2003}, the PDR has
a density gradient between the star and the molecular cloud. For the present
preliminary model, we represent this interface with two
effective media at low (200 cm$^{-3}$) and high ($2 \times 10^4$~cm$^{-3}$)
densities irradiated by strong and weak radiation fields. 
In the more excited $\rho$~Ophiuchi region we are able to match the AME by
requiring: (a) an attenuated radiation field in the dense gas; and (b)
smaller and less abundant PAHs in stronger radiation
fields, in line with recent
work suggesting that PAHs are photoevaporated from larger carbonaceous
grains \citep{Berne2007,Compiegne2008,Velusamy2008}. The PAH
abundances we assume for both regions are within the range found by \cite{Habart2003}. 
The fit is shown in Fig.~\ref{fig:roph_residual} and 
the parameters of our model are summarised in
Table~\ref{tab:model_param}.

\begin{figure*}
\centering
\includegraphics[width=1.0\textwidth]{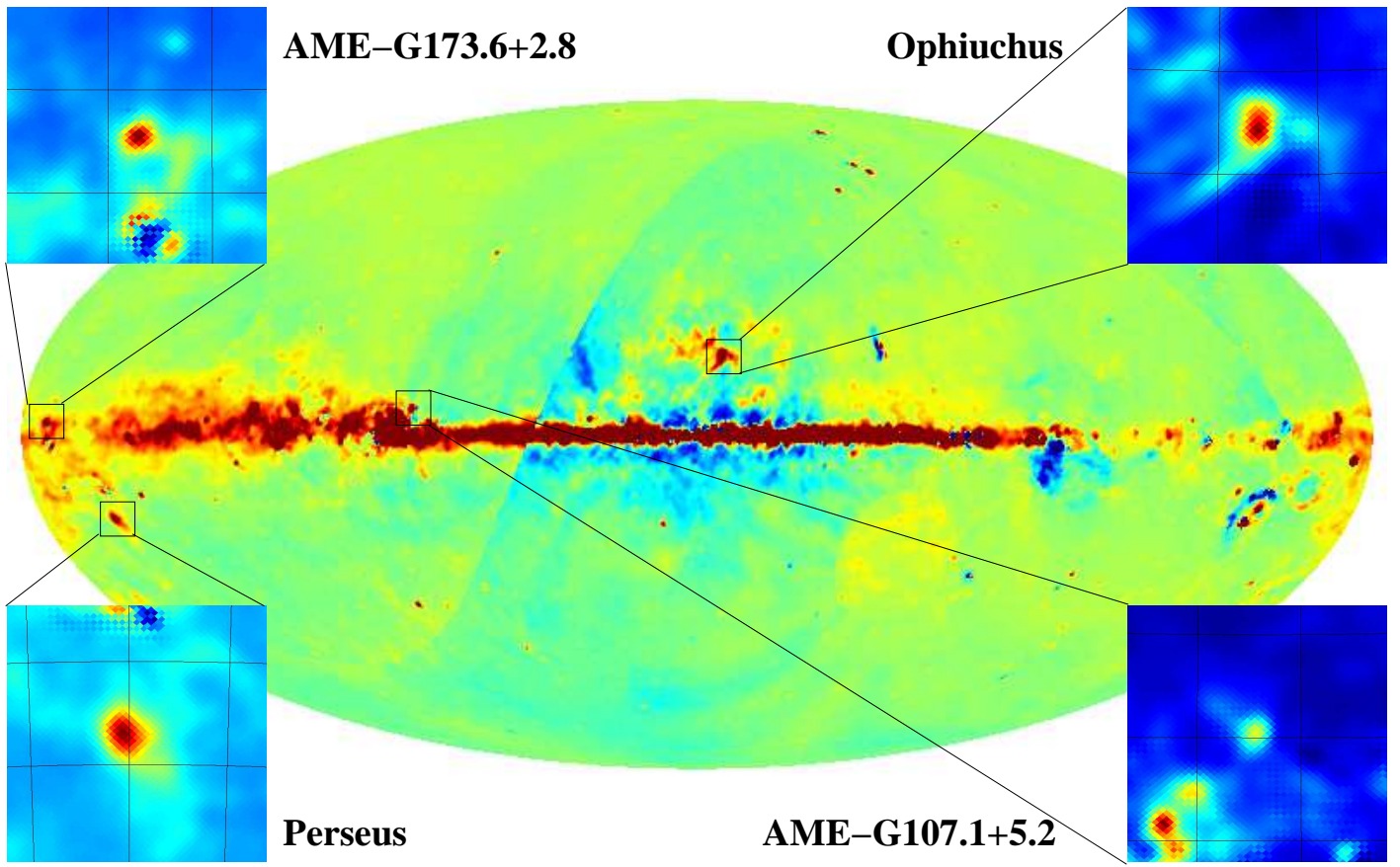}
\caption{Residuals in the full sky \planck~LFI 28.5\,GHz $1\degr$ smoothed map after subtraction of synchrotron, free-free and thermal dust emission (see text). $12\fdg5 \times 12\fdg5$ cut out maps are shown for the Perseus and $\rho$ Ophiuchi molecular clouds, and the two new regions of AME, AME-G107.1+5.2 and AME-G173.6+2.8. The graticule spacing is $5\degr$ in Galactic coordinates.}
\label{fig:planck30subtracted}
\end{figure*}

We note, however, that the determination of $a_0$, $G_0$ and $n_{\rm H}$ is degenerate in the
present modelling; you can create almost identical spinning dust curves by varying these parameters in different ways (see, e.g., \citealt{Ali-Hamoud2009}). Lifting this degeneracy involves treating the
radiative transfer along with the spinning dust motion. The present results are therefore suggestive only and we emphasise that the
physical conditions of our models are derived from scales much smaller
than the resolution of our observations. The previous discussion
also indicates that quantitative model spectra can only be
obtained from a consistent treatment of the gas state, the radiative
transfer, and the spinning motion of dust grains. Finally, a combination of multi-frequency data, including detailed
IR measurements, is clearly needed to extract detailed information
about the environment and the dust grains.

%%%%%%%%%%%%%%%%%%%%%%%%%%%%%%%%%%%%%%%%%%%%%%%%%%%%%%%%%%%%%%%%%%%%%%%%%%%%%%%%%

\section{New regions of anomalous emission}
\label{sec:new_regions}

We have searched the \Planck\ maps for new regions of AME on the scale of $\sim\negmedspace3\degr$ or smaller. To do this, one must suppress the other emission components that contribute to the frequencies where AME is the strongest ($\sim$20--60\,GHz). Numerous methods can be used (e.g., \citealt{Leach2008,Delabrouille2009}). However, we have taken a simplistic approach to identify bright AME regions. We subtracted a simple spatial model of known emission mechanisms (synchrotron, free-free, and thermal dust) using extrapolations of existing templates from observations or theoretical expectations, and then examined the locations of the remaining emission in more detail. We used maps smoothed to $1\degr$, with HEALPix $N_{\rm side}=256$ \citep{Gorski2005}.

We subtracted the synchrotron emission from the CMB-subtracted \Planck\ maps using the 408\,MHz map \citep{Haslam1982}, while the 1420~\,MHz \citep{Reich1982,Reich1986,Reich2001} and the 2326\,MHz \citep{Jonas1998} maps were used in conjunction with 408\,MHz data to determine the spectral index. Brightness temperature spectral indices were restricted to lie in the range $-2.0 > \beta > -4.0$ ($T \propto \nu^{\,\beta}$).  For the gap region where no 1420/2326\,MHz data exist, we assume $\beta=-3.0$. We used versions of the maps that have had no point sources subtracted \citep{Platania2003} so as to allow direct comparison with the \Planck~data. For free-free emission, we use the H$\alpha$ map of \cite{Dickinson2003}, which is based on WHAM \citep{Haffner2003} and SHASSA \citep{Gaustad2001} data corrected for dust absorption. The map is also used to correct the synchrotron maps for free-free emission, assuming an average electron temperature of $8000$\,K. The Rayleigh-Jeans tail of the thermal (vibrational) dust component is determined from the \Planck~maps at 143 and 545\,GHz, thereby avoiding significant CO contamination. The 545\,GHz data were used for normalisation and the 143\,GHz data were used for the spectral index of the power-law. This was constrained to lie between $0$ and $+3$ to avoid extrapolation artefacts in some areas at high latitudes where the uncertainty on the spectral index is large.

The full-sky map of the residuals from the 28.5\,GHz \Planck~LFI data is shown in Fig.~\ref{fig:planck30subtracted}. Artefacts from the subtraction are present. The division between the 1420 and the 2326~MHz maps is evident as well as the lack of observations around the south celestial pole. Other large-scale features are also present. These are extended features and hence do not affect our search for compact areas of AME. A large number of AME candidates are evident in the map, including the strong Perseus and $\rho$~Ophiuchi AME regions described in Sects.~\ref{sec:results_perseus} and \ref{sec:results_roph}. We examined $\sim 50$ regions in detail.

The next step in identifying new AME regions is to construct a spectrum for each candidate, using aperture photometry. In each case there is strong thermal dust and free-free emission indicative of an associated \ion{H}{ii} region. We now illustrate the results of our search in two of the compact regions which show excess emission in the $\sim$20--60\,GHz range.

\subsection{AME-G173.6+2.8}
AME-G173.6+2.8 ($05^\mathrm{h}41^\mathrm{m}, +35^\circ51\arcmin$ J2000.0) is a strong, compact region of dust emission towards the Galactic anticentre.  The \ion{H}{ii} region S235 \citep{Sharpless1959} lies within this area. It is a region of massive star formation with a young stellar object at its centre, and it has several compact radio components within several arc\-min \citep{Nordh1984,Felli2006}. Radio recombination line observations indicate a radial velocity of $-24.5 \pm 0.7~$km~s$^{-1}$ \citep{Paladini2003}. S235 lies at a distance of 1.8\,kpc and is therefore within the Perseus spiral arm.

\begin{figure}
\centering
\includegraphics[width=0.5\textwidth]{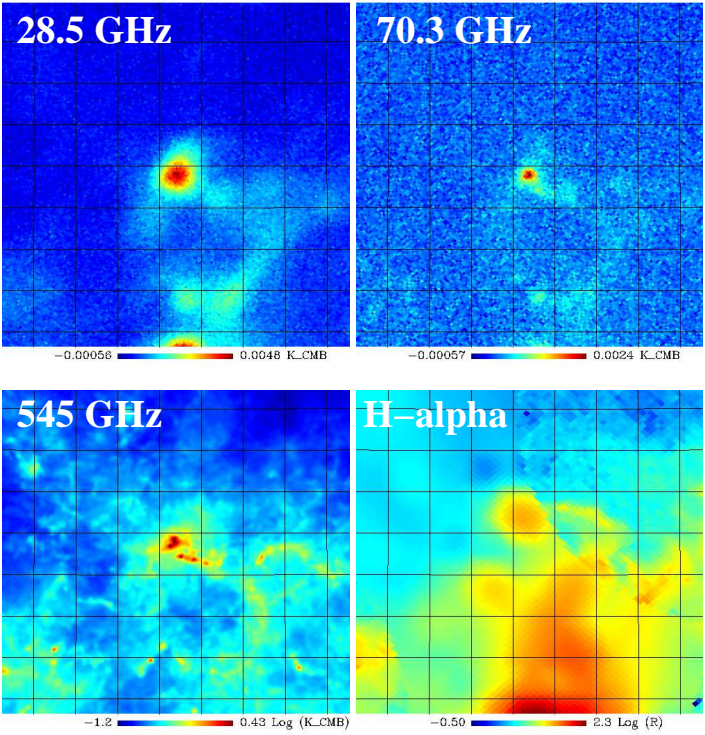}
\caption{Maps of the AME-G173.6+2.8 region: 28.5\,GHz (linear scale); 70.3\,GHz (linear scale); 545\,GHz (logarithmic scale); and H$\alpha$ (logarithmic scale). The maps are $8\degr \times 8\degr$ and the graticule spacing is $1\degr$ in Galactic coordinates.}
\label{fig:g173_maps}
\end{figure}

Maps of the region at a range of frequencies are shown at the observing resolution of each in Fig.~\ref{fig:g173_maps}. The AME is strong at 28.5\,GHz while the dust distribution at the higher \Planck~frequencies shows the small-scale structure of the dust within the 28.5\,GHz beam. It can be seen that the H$\alpha$ at the $6\arcmin$ resolution of \cite{Finkbeiner2003} is offset from the dust, most likely due to absorption by the dust.

\begin{figure}
\centering
\includegraphics[width=0.35\textwidth,angle=-90]{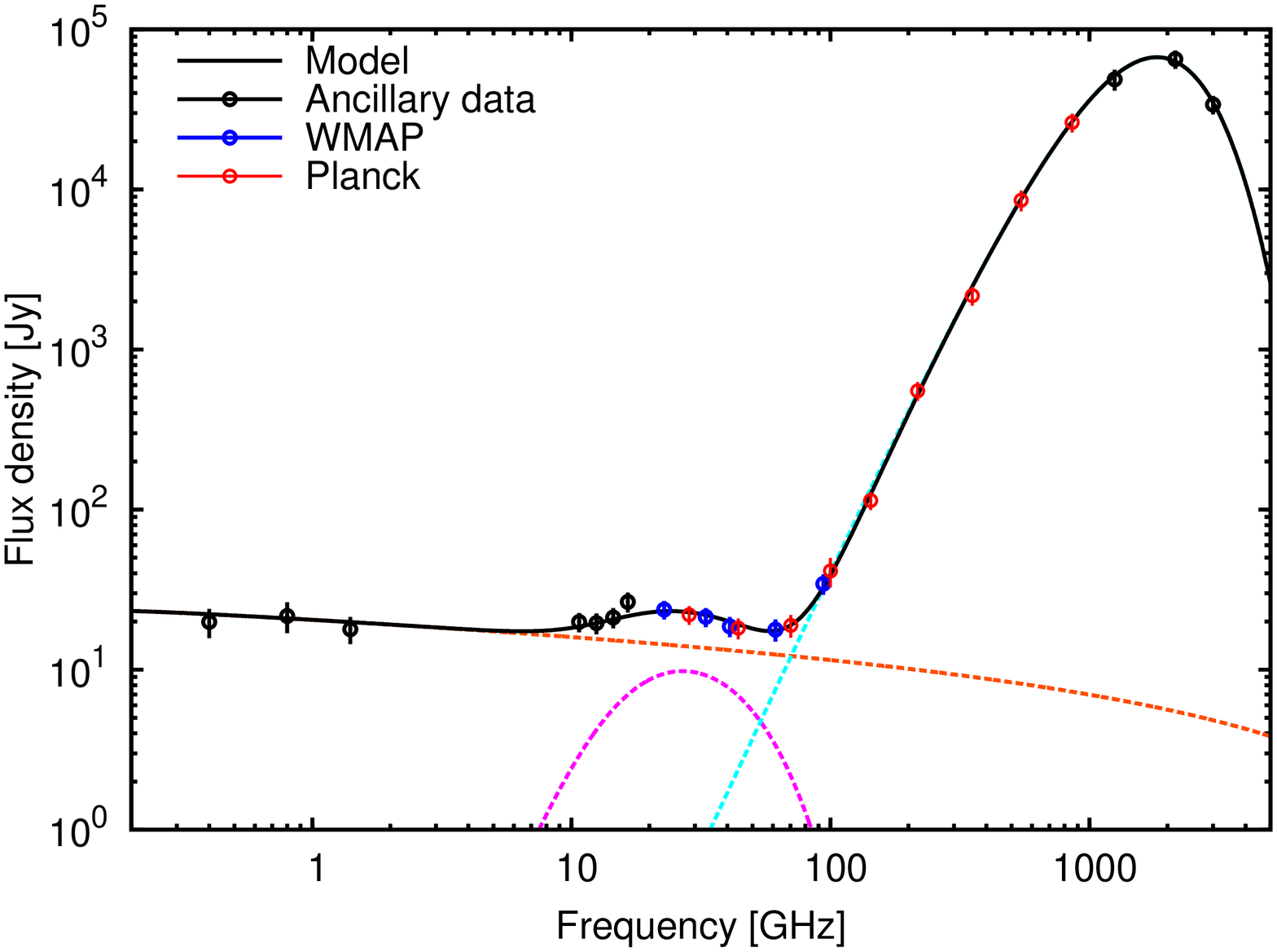}
\includegraphics[width=0.35\textwidth,angle=-90]{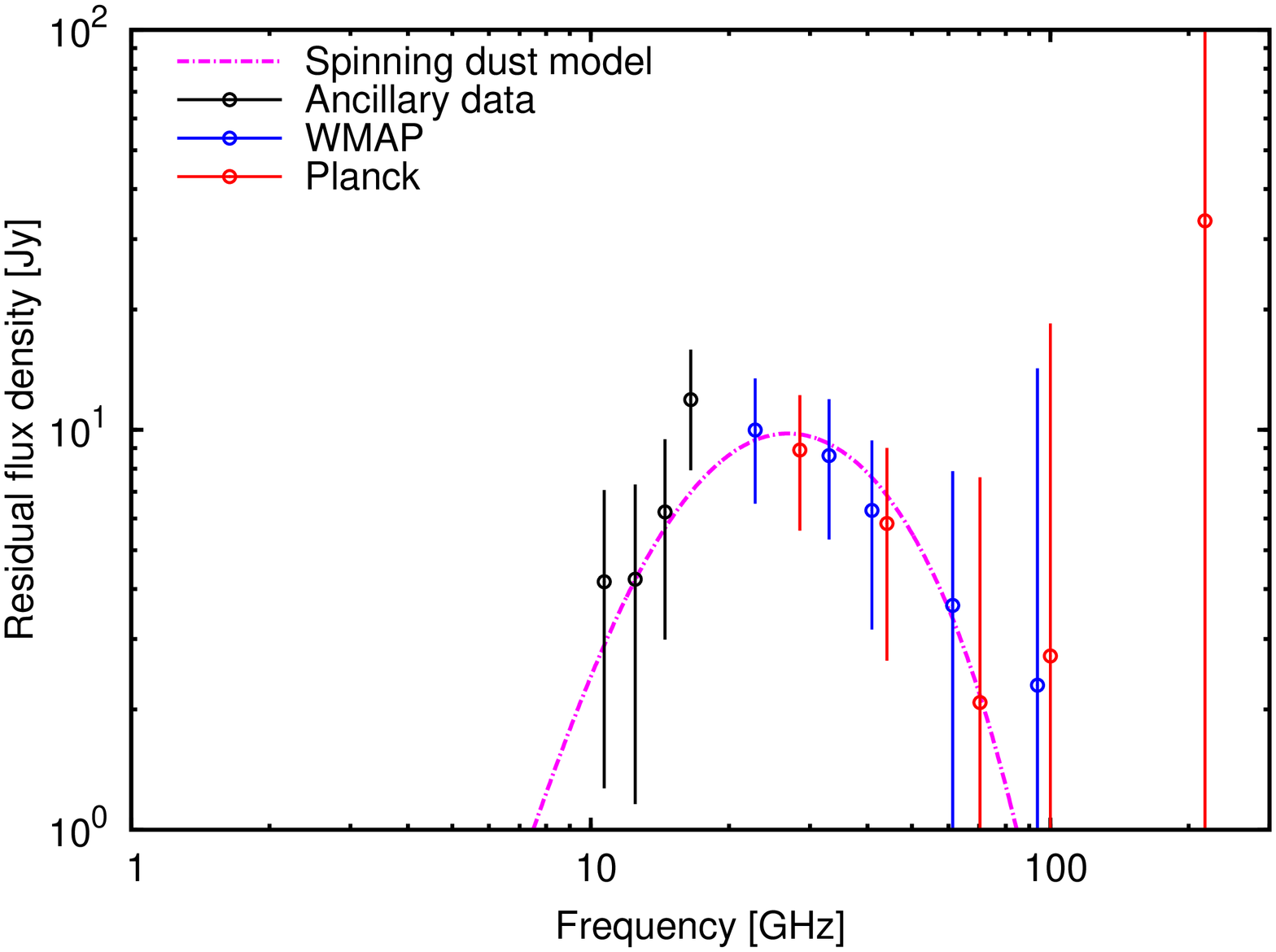}
\caption{The spectrum (top) and residuals (bottom) for AME-G173.6+2.8 after removing free-free emission (orange dashed line), thermal dust emission (light blue dashed line) and CMB anisotropies (not visible). A spinning dust model is shown as a magenta dot-dashed line.}
\label{fig:g173_sed}
\end{figure}

The spectrum of the $1\fdg3$ degree (COSMOSOMAS) region centred on AME-G173.6+2.8 is shown in Fig.~\ref{fig:g173_sed}. The radio flux densities are at 820\,MHz \citep{Berkhuijsen1972}, 1.4\,GHz \citep{Reich1982} and $11$--$17$\,GHz (COSMOSOMAS). The spectrum and residuals of the region are derived using the COSMOSOMAS filtering method described in Sect.\,\ref{sec:perseus_spectrum}.

The spectrum is well-fitted using a combination of thermal dust (with $T = 19.6\pm0.6$\,K and an emissivity index of $\beta=1.54\pm0.07$), free-free emission (${\rm EM}=(2.47\pm0.13)\times10^{3}$\,cm$^{-6}$\,pc) and spinning dust (\cite{Draine98b} CNM model with $N_{\rm H}=(2.7\pm0.3) \times 10^{22}$\,cm$^{-2}$), with a small CMB contribution of $\Delta T_{\rm CMB}=130\pm40\,\upmu$K. The residual spectrum shows the signature of spinning dust peaking in the frequency range $20$--$30$\,GHz, detected with a significance level of $6.4\sigma$. The main uncertainty in the AME spectrum is the level of free-free emission to be subtracted. As shown, the spinning dust emission in the 20--30\,GHz region is $\sim 2$ times the free-free emission; the uncertainty is $<10$--$20\,\%$. The current spectral coverage defines the peak of the spectrum and provides useful constraints in modelling the AME. An ultracompact \ion{H}{ii} region, producing free-free emission that is optically thick at frequencies $<$10\,GHz, may be contributing to the flux density at $>$10\,GHz. However, it cannot easily account for the shape at high ($\gtrsim 40$\,GHz) frequencies.

\subsection{AME-G107.1+5.2}
AME-G107.1+5.2 ($22^\mathrm{h}22^\mathrm{m}, +63^\circ23\arcmin$ J2000.0) is a dust complex containing the star-forming \ion{H}{ii} region S140 \citep{Sharpless1959} with $V_{\rm LSR}=-8.5\pm1.0~$km~s$^{-1}$. The region  lies at the edge of the Cepheus bubble \citep{Patel1998} at a distance of 800\,pc. 

\begin{figure}
\centering
\includegraphics[width=0.5\textwidth]{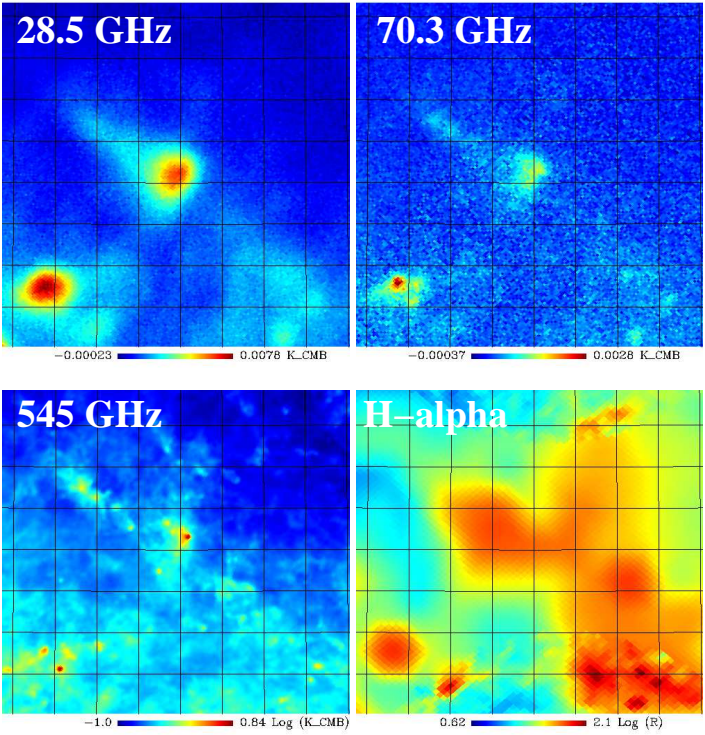}
\caption{Maps of the AME-G107.1+5.2 region: 28.5\,GHz (linear scale); 70.3\,GHz (linear scale); 545\,GHz (logarithmic scale); and H$\alpha$ (logarithmic scale). The maps are $8\degr \times 8\degr$ and the graticule spacing is $1\degr$ in Galactic coordinates.}
\label{fig:g107_maps}
\end{figure}

Maps of the region are shown in Fig.~\ref{fig:g107_maps} using the same format as in Fig.~\ref{fig:g173_maps}. Dust is widespread over the region but with a concentration at the position of the AME. There are multiple dust components within the 28.5\,GHz beam. There is a small offset of the 28.5\,GHz feature of $\approx 15 \arcmin$ to lower latitudes and higher longitudes than the brightest feature in the dust maps. This indicates a higher radio emissivity of the dust in this offset position. There is also an offset in the H$\alpha$ emission, again possibly due to absorption by the associated dust.

\begin{figure}
\centering
\includegraphics[width=0.35\textwidth,angle=-90]{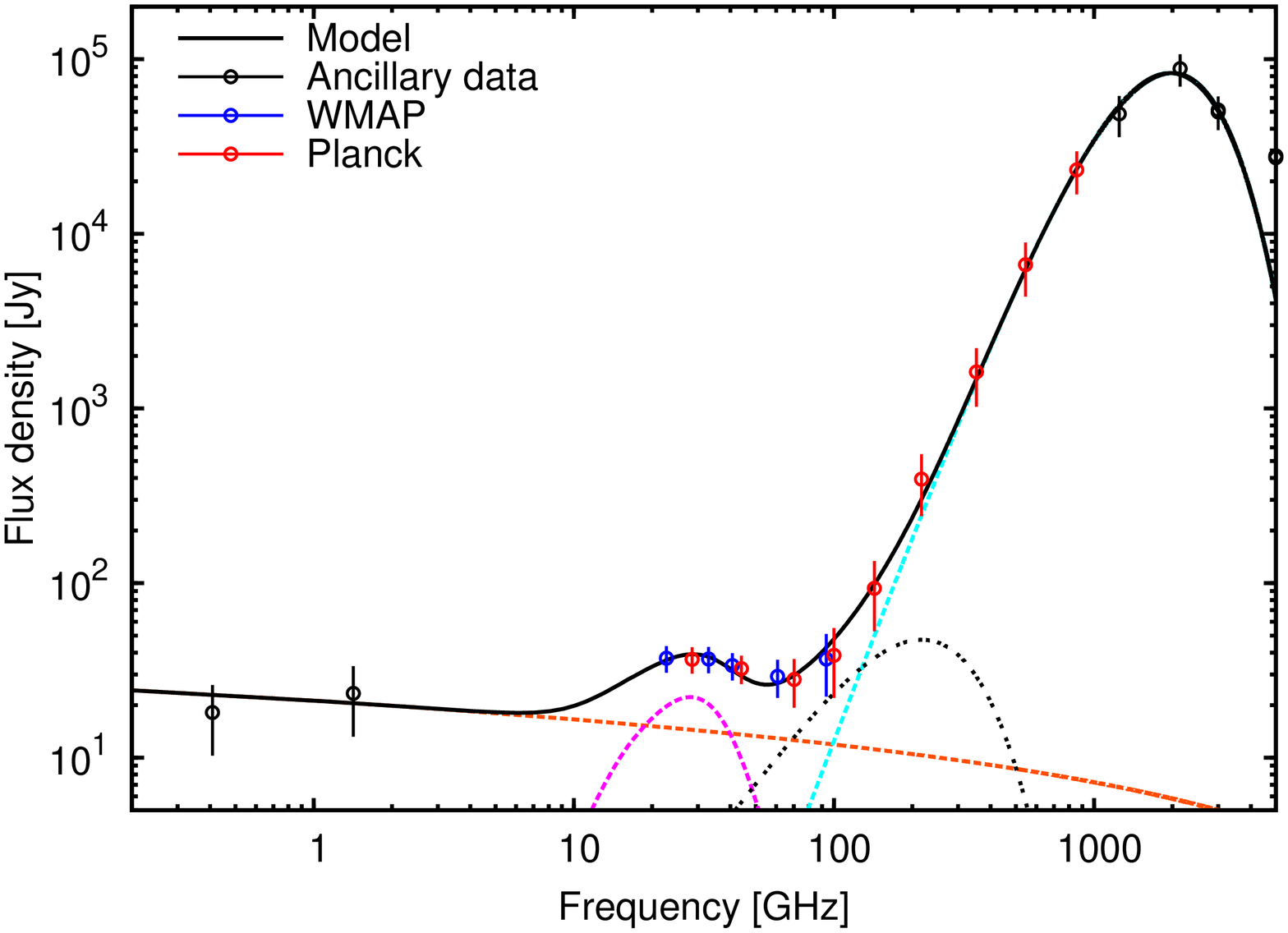}
\includegraphics[width=0.35\textwidth,angle=-90]{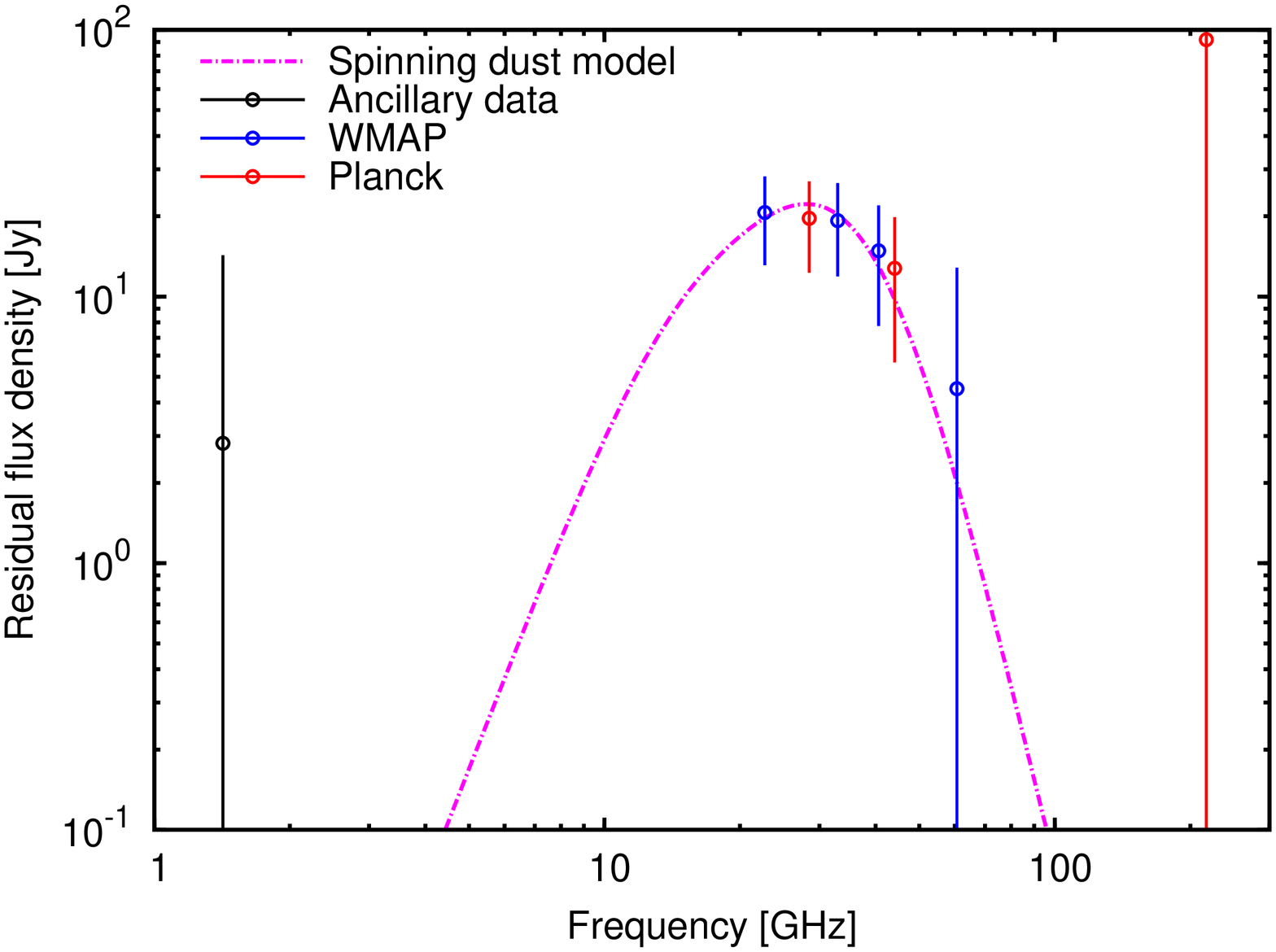}
\caption{The spectrum (top) and residuals (bottom) for AME-G107.1+5.2 after removing free-free (orange dashed line), thermal dust emission (light blue dashed line) and CMB anisotropies (black double-dotted line). A spinning dust model is shown as a magenta dot-dashed line.}
\label{fig:g107_sed}
\end{figure}

The spectrum estimated for AME-G107.1+5.2 and the residual spectrum are shown in Fig.~\ref{fig:g107_sed}. The least-squares fit to the free-free emission (${\rm EM}=150\pm39$~cm$^{-6}$~pc), thermal dust ($T_{\rm d}=19.0\pm2.6$~K, $\beta=2.04\pm0.45$), spinning dust (\cite{Draine98b} WIM model with $N_{\rm H}=(1.76 \pm 0.44) \times 10^{21}$~cm$^{-2}$), and CMB ($\Delta T_{\rm CMB}=75\pm38\,\upmu$K) is well-defined. The AME is detected at a significance of $4.8\sigma$ with a shape that is consistent with spinning dust. However, an ultracompact \ion{H}{ii} region producing optically thick free-free at frequencies below 10\,GHz may also be contributing. 

\subsection{Assessment of the search of new AME regions}

We have shown that it is possible to detect bright compact AME regions in the \Planck~data using a rather simplistic selection process.  In order to verify the identifications it is necessary to add ancillary data at radio and FIR frequencies.  Now that identifications have been made, new ancillary data covering a larger frequency range will be sought. Further development of the search algorithms will also be worthwhile.     

The strength of AME relative to the free-free and thermal dust emission in these two regions can be compared with the values for \ion{H}{ii} regions obtained by \cite{Todorovic2010} in the VSA survey of the $l=27\degr$--$47\degr$ section of the Galactic plane.  The latter survey found the ratio of 33\,GHz AME to  free-free flux density to be $0.40\pm0.11$; the corresponding values for AME-G173.6+2.8 and AME-G107.1+5.2 are 0.67 and 1.4, respectively.  In the case of the ratio of 33\,GHz~AME to $100\,\upmu$m thermal dust emission, the VSA survey value was $1.1 \times 10^{-4}$ (or $2\times 10^{-4}$ at the 15\,GHz peak of the spectrum), while the AME-G173.6+2.8 and the AME-G107.1+5.2 values are somewhat larger at $3.3 \times 10^{-4}$ and $3.8 \times 10^{-4}$, respectively. It is however not surprising that the 30\,GHz AME is stronger in the two \Planck\ \ion{H}{ii} regions than in the VSA \ion{H}{ii} regions, because the former were selected as the brightest AME sources in the \Planck~maps. The contribution from ultracompact \ion{H}{ii} regions \citep{Wood1989} may also be a significant contributor to the AME at low Galactic latitudes.

Finally, we note that the derived dust temperature for these \ion{H}{ii} regions, at $T_{\rm d}\approx 19$\,K, is closer to that of the diffuse interstellar medium than to the warm dust ($T_{\rm d} \sim 30$--$50$~K) usually found surrounding OB stars (e.g., \citealt{Wood1989}). The dust observed along the line of sight at $\sim 1\degr$ resolution is likely to be a mixture of populations, including higher density regions (e.g., molecular clouds) that could dominate the signal on these scales. This may explain the lower average temperature the presence  of significant AME in these sightlines.

As selection procedures develop, we expect many AME regions to be discovered in a range of physical environments, which will provide data necessary to acquire a full understanding of the AME emission mechanism. 

%%%%%%%%%%%%%%%%%%%%%%%%%%%%%%%%%%%%%%%%%%%%%%%%%%%%%%%%%%%%%%%%%%%%%%%%%%%%%%%%%

\section{Conclusions}
\label{sec:conclusions}

Filling in the gap between earlier radio and FIR measurements, the
frequency coverage of \Planck\ has provided a unique opportunity to
establish comprehensive spectra of regions of AME. \Planck~has
revealed the high frequency side of the spectral peak. The new spectra
are the basis for understanding the emission mechanism and the
environment in which it occurs. The evidence from the present
observations strongly favours the spinning dust mechanism (electric
dipole radiation). \Planck\  provides a rich data set that can be used as
a basis for developing a realistic understanding of the AME mechanism
in a range of Galactic environments.

The two best-studied AME sources that have extensive ancillary data
are the Perseus and $\rho$ Ophiuchi molecular clouds. In each, the
spectrum is well fitted by free-free, thermal dust, and spinning dust,
with a small contribution from the CMB. Spinning dust provides a good
fit to the microwave ($10$--$100$\,GHz) part of the spectrum, which
peaks at $\approx 30$\,GHz.

Theoretical spinning dust spectra are presented for a physical model
consisting of molecular and atomic states. It is possible to derive
physical parameters that are consistent with the environment and still
provide a good fit to the data. Using parameters constrained at
smaller angular scales, the $20$--$40$\,GHz AME peak in Perseus is
well explained with spinning dust emission arising from dense,
molecular gas ($n_{\rm H}>200$~cm$^{-3}$) subjected to a few times the
typical interstellar radiation field. Low-density gas is only a minor
contributor to AME in  Perseus; however, in $\rho$ Ophiuchi, although dense gas accounts for the peak at
$\approx 30$\,GHz, irradiated low-density atomic gas may be
contributing in the range $50$--$100$\,GHz. The picture seems to be
that smaller PAHs are found in PDRs  ($G_0>100$), as suggested by
recent {\it Spitzer} observations. However, the determination of the
PAH size is degenerate with that of $n_{\rm H}$ and $G_0$ and
quantitative conclusions will only be obtained from consistent
modelling of the gas state, radiative transfer, and spinning dust. At
this level of modelling it is not possible to constrain the electric
dipole moment of PAHs. Future work to carry out more detailed
modelling is required.

Our preliminary search for new AME regions in the \Planck~data has
revealed many candidates. They were uncovered by subtracting
synchrotron, free-free and thermal dust emission based on the usual
spatial templates. Two of the new candidate regions that show AME at
$>5\sigma$ significance have spectra consistent with a spinning
dust spectral shape. Additional multi-frequency data (e.g., such as the 5\,GHz C-Band All-Sky Survey, C-BASS\footnote{http://www.astro.caltech.edu/cbass/}; \citealt{King2010}) as well as high-resolution observations will
be needed to understand the detailed structure of the AME in these
regions.

%%%%%%%%%%%%%%%%%%%%%%%%%%%%%%%%%%%%%%%%%%%%%%%%%%%%%%%%%%%%%%%%%%%%%%%%%%%%%%%%%

\begin{acknowledgements}
We thank the referee, Doug Finkbeiner, for useful comments. We thank Justin Jonas for providing the 2326~MHz HartRAO map. We acknowledge the use of the MPIfR Survey Sampler website at \url{http://www.mpifr-bonn.mpg.de/survey.html}. We acknowledge the use of the Legacy Archive for Microwave Background Data Analysis (LAMBDA); support for LAMBDA is provided by the NASA Office of Space Science. This research has made use of the NASA/IPAC Extragalactic Database (NED) which is operated by the Jet Propulsion Laboratory, California Institute of Technology, under contract with the National Aeronautics and Space Administration. This research makes use of the SIMBAD database, operated at CDS, Strasbourg, France.

The Planck Collaboration acknowledges the support of: ESA; CNES and CNRS/INSU-IN2P3-INP (France); ASI, CNR, and INAF (Italy); NASA and DoE (USA); STFC and UKSA (UK); CSIC, MICINN and JA (Spain); Tekes, AoF and CSC (Finland); DLR and MPG (Germany); CSA (Canada); DTU Space (Denmark); SER/SSO (Switzerland); RCN (Norway); SFI (Ireland); FCT/MCTES (Portugal); and DEISA (EU). A detailed description of the Planck Collaboration and a list of its members can be found at \url{http://www.rssd.esa.int/index.php?project=PLANCK&page=Planck_Collaboration}. 

\end{acknowledgements}

\bibliographystyle{aa}

\bibliography{Planck_bib,clive_refs}
\raggedright 
\end{document}